\documentclass[a4paper,amsfonts,amssymb,amsmath,reprint,nofootinbib,twoside,superscriptaddress,floatfix]{revtex4-2}
\usepackage[english]{babel}
\usepackage[utf8]{inputenc}
\usepackage[T1]{fontenc}
\usepackage{lmodern,microtype,mathtools,bm,graphicx,xcolor,booktabs}
\usepackage[notrig]{physics}
\usepackage{amsthm}
\usepackage{hyperref}
\hypersetup{colorlinks=true,citecolor=blue!55!black,linkcolor=blue!55!black,urlcolor=blue!55!black}

\newtheorem{theorem}{Theorem}
\newtheorem{proposition}{Proposition}
\renewcommand{\theproposition}{S.\arabic{proposition}}
\newcommand{\C}{\mathbb C}
\newcommand{\T}{\mathrm{\mathbb{T}}}
\newcommand{\diag}{\operatorname{diag}}
\newcommand{\perm}{\operatorname{perm}}
\providecommand{\Real}{\operatorname{Re}}
\providecommand{\Imag}{\operatorname{Im}}
\providecommand{\F}{\mathbb F}

\providecommand{\figlink}[2]{\hyperref[#1]{\textcolor{blue}{#2}}}
\begin{document}
\raggedbottom

\title{Using topology to reveal all mutually unbiased bases in dimension six}

\author{Mateo C\'ardenes Wuttig}
\email{mateo.cardeneswuttig@yale.edu}
\affiliation{Department of Applied Physics, Yale University, New Haven, Connecticut 06520, USA}
\affiliation{Yale Quantum Institute, Yale University, New Haven, Connecticut 06520, USA}
\author{Joseph Tindall}
\affiliation{Center for Computational Quantum Physics, The Flatiron Institute, 162 5th Avenue, New York, New York 10010, USA}
\date{\today}

\begin{abstract}
Mutually unbiased bases (MUBs) describe complementary quantum measurements: certainty in one basis gives no information about the others. In dimension six, triples of MUBs are known, but whether a fourth basis could exist remained unresolved for over three decades. OpenAI recently reported a proof that none exists. Here we use the topology of the space of complex Hadamard matrices, which describe transitions between MUBs, to characterize all sets in dimension six. We prove that every mutually unbiased triple belongs to a two-parameter family or is a single isolated triple built from the Fourier matrix. Since none of them extend to a fourth basis, this also proves that no four MUBs exist. Our proof rests on a new topological result that all order-six complex Hadamard matrices except Tao's form a connected space. Connectedness reduces the classification of triples to finitely many overlapping four-dimensional phase balls. We certify with interval arithmetic that these balls cover the space and for every pair of MUBs represented in a ball, we then determine all third bases that extend it. Dimension six is now the first non-prime-power dimension in which all sets of MUBs are classified.
\end{abstract}

\maketitle

\section{Introduction}

\begin{figure*}[htb!]
 \centering
 \includegraphics[width=\textwidth]{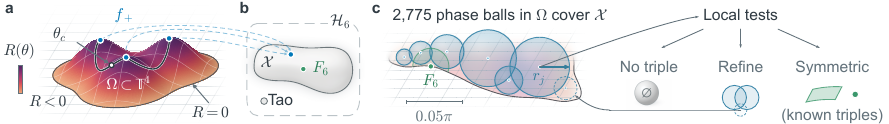}
\caption{\textbf{Schematic structure of our topological proof, which classifies all MUBs in dimension six.}
\textbf{a,} Four phases $\theta$ fix a dephased $3\times3$ corner, which can be completed to a Hadamard matrix when the reconstruction invariant $R(\theta)$ is positive. These \textit{strict} seeds form the region $\Omega=\{\theta:R(\theta)>0\}$, bounded by $R=0$. Explicit paths such as $\theta_c$ (see Eq.~\ref{eq:omega-joining-circle}) join all local maxima of $R$ (blue points), so $\Omega$ is connected.
\textbf{b,} The continuous reconstruction $f_+(\theta)$ maps $\Omega$ to Hadamard classes, sending all maxima to one class (blue point). Adding the Fourier matrix $F_6$ (green point) gives the connected space $\mathcal X$ of all non-Tao Hadamard matrices.
\textbf{c,} We cover $\mathcal X$ with the image of $2{,}775$ overlapping phase balls, each certified to lie strictly inside $\Omega$, together with a separate chart for $F_6$ (green). In each ball, we rule out a triple through a graph search for six mutually orthogonal candidate vectors and local exclusion tests, or we prove that every triple has an order-three symmetry. These symmetric triples are exactly the known types: the two-parameter family (green patch) and the isolated Fourier triple (green point). This approach covers all MUB sets in dimension six, and since the known triples cannot be extended to a quartet, no four MUBs exist.}
 \label{fig:sketch}
\end{figure*}

How many quantum measurements can be complementary, and what forms do they take?
Mutually unbiased bases (MUBs) answer this question mathematically. Two orthonormal bases $\mathcal A=\{\ket{a_i}\}$ and $\mathcal B=\{\ket{b_j}\}$ in $\mathbb C^d$ are \textit{mutually unbiased} when
\begin{equation}
 \abs{\braket{a_i}{b_j}}^2=\frac1d
 \quad\text{for all }i,j.
 \label{eq:mub}
\end{equation}
Preparing a state in one basis then gives equal probabilities for every outcome of the other measurement basis. MUBs connect this form of complementarity~\cite{Schwinger1960,Ivanovic1981} to state reconstruction~\cite{WoottersFields1989}, uncertainty relations~\cite{MaassenUffink1988}, quantum key distribution~\cite{Cerf2002}, and entanglement tests~\cite{Spengler2012}.
They are closely tied to complex Hadamard matrices, which describe the transitions between different MUBs~\cite{Durt2010,TadejZyczkowski2006}.

The largest number of MUBs, $N(d)$, determines how many perfectly complementary measurements a $d$-level quantum system supports. At most $d+1$ are possible, and \emph{complete sets} attaining this bound allow full quantum state reconstruction and exist whenever $d$ is prime or a power of a prime~\cite{WoottersFields1989,Bandyopadhyay2002,Klappenecker2004,Weiner2013}. 
Dimension six is the smallest dimension that is not a prime power. Tensor products of MUBs in dimensions two and three give three bases in $\mathbb C^6\cong\mathbb C^2\otimes\mathbb C^3$~\cite{Klappenecker2004}, which have been realized with single photons using polarization and three orbital-angular-momentum modes~\cite{DAmbrosio2013}. 
In 1991, Zauner conjectured that three MUBs is the maximum in dimension six~\cite{Zauner1991,Zauner1999}, a question that remained open for decades~\cite{KostrikinEtAl1981,WoottersFields1989Searching,Zauner1991,Zauner1999,KruegerWerner2005,HorodeckiRudnickiZyczkowski2022}.

Most attempts to resolve this conjecture start from a pair of bases, represented by a $6\times6$ complex Hadamard matrix, and ask whether it is compatible with a third or fourth basis. Many families of such matrices are known~\cite{BengtssonEtAl2007,BeauchampNicoara2008,MatolcsiSzollosi2008,SkinnerNewellSanchez2009,Szollosi2010Hypocycloids,Karlsson2011ThreeParameter,BarrosSaBengtsson2013,BondalZhdanovskiy2016}, and several have been excluded from participating in three or four MUBs. Grassl ruled out quartets containing the standard and Fourier bases~\cite{Grassl2004}, and Jaming \emph{et al.} extended this to the whole two-parameter Fourier family~\cite{Jaming2009}. Tao's isolated matrix does not even admit a third basis~\cite{BrierleyWeigert2009}, and triples of product bases admit no fourth~\cite{McNultyWeigert2012AllProduct,McNultyWeigert2012Triples,McNultyWeigert2012LimitedRole}. Further work has constrained the structure any quartet must have~\cite{AschbacherChildsWocjan2007,BoykinEtAl2007,ChenYu2017,McNultyWeigert2025Comment,LiangChen2025Reply,Goyeneche2013ER,HuangLiangChen2026ThreeElements}, and a wide range of analytical and numerical approaches has been explored~\cite{JamingMatolcsiMora2010,Matolcsi2012,MaxwellBrierley2015,MatolcsiWeiner2015,KolountzakisMatolcsiWeiner2018,BandeiraDoppelbauerKunisky2022,McConnell2021,ReimpellWerner2007,ButterleyHall2007,RaynalLuEnglert2011,Goyeneche2013,BatleEtAl2016,GultekinEtAl2026,FatahEtAl2026,FuhrRzeszotnik2015,PratColomer2022,AguilarEtAl2018QRAC,BrierleyWeigert2010SDP,GriblingPolak2024,Mortimer2023}, including for smaller sets of mutually unbiased vectors known as \emph{constellations}~\cite{BrierleyWeigert2008}. 

None of these results rule out a fourth basis in general~\cite{McNultyWeigert2026}. However, every known triple belongs to one of two types, a two-parameter family and a single isolated triple built from the Fourier matrix~\cite{Goyeneche2013,MatszangoszSzollosi2024}. Both cannot be extended to four MUBs because they are related by a matrix in the Fourier family~\cite{Jaming2009}. Matolcsi, Matszangosz, Varga and Weiner searched numerically for triples, found only these two types, and conjectured that no others exist~\cite{MatolcsiEtAl2026MUBTriplets}, which would also imply Zauner's conjecture that no four MUBs exist~\cite{Zauner1991,Zauner1999}.

Recent months have seen rapid progress on mutually unbiased bases, much of it assisted by AI~\cite{wuttig2026completeclassificationcomplexhadamard,wuttig2026mutuallyunbiasedbases,Sun2026Completion,ranchin2026newnonaffinefamilykarlssonszollosi}. It was recently proved that every order-six complex Hadamard matrix can be reconstructed algebraically from 
just four phases~\cite{wuttig2026completeclassificationcomplexhadamard}, building on Sz\"oll\H{o}si's original construction~\cite{Szollosi2012FourParameter}, and larger sets of MUBs than previously known were constructed~\cite{wuttig2026mutuallyunbiasedbases}, including five in dimension twelve, where earlier numerical searches had failed~\cite{WuLiuWu2026}. Most recently, OpenAI reported a computer-assisted proof that at most three MUBs exist in dimension six~\cite{OAIMUB2026}.
Their proof covers all Hadamard matrices with balls around nine phases and uses floating-point interval bounds and a graph search inside each ball to exclude a fourth basis. It does not use results on the known Hadamard families and leaves the triplet classification question open.

Here, we classify all sets of MUBs in dimension six up to standard equivalence. Mutually unbiased pairs are given by our reconstruction of order-six complex Hadamard matrices, including pairs that admit no third basis. For triples, we prove the conjecture of Matolcsi, Matszangosz, Varga and Weiner~\cite{MatolcsiEtAl2026MUBTriplets}: every triple is equivalent either to a member of the two-parameter family or to the isolated Fourier triple~\cite{Goyeneche2013,MatszangoszSzollosi2024}, which also proves that no four MUBs can exist~\cite{Zauner1991,Zauner1999,Jaming2009}.
Dimension six thus becomes the first non-prime-power dimension in which all sets of MUBs are classified.

Our proof, sketched in Fig.~\ref{fig:sketch}, rests on the topology of the space of complex Hadamards. First, we simplify our earlier reconstruction~\cite{wuttig2026completeclassificationcomplexhadamard} and show that every order-six complex Hadamard matrix except Tao's can be built from four phases, with the Fourier matrix $F_6$ reached as a limit. We then prove that the region of phases on which this reconstruction applies is topologically connected, and so is the space of all non-Tao Hadamard matrices.
By connectedness, a finite union of closed balls is the whole space as soon as every boundary point of each ball lies in the interior of another. We use $2{,}775$ balls in the four phase coordinates, plus one additional chart around $F_6$, and use interval arithmetic to verify that every triple whose second basis lies in a ball is of a known type. Since Tao's matrix admits no third basis~\cite{BrierleyWeigert2009}, this completes the classification of all triples. Finally, we show that our results also give an analytical proof of the Fourier-vanishing conjecture of Matolcsi, Ruzsa and Weiner~\cite{MatolcsiRuzsaWeiner2013}, for which OpenAI recently reported a computer-assisted proof~\cite{OAIFourier2026}. 

The reconstruction and the connectedness of the space are formalized in Lean~\cite{deMouraUllrich2021Lean4}. The cover and the exclusion of quartets in each ball are verified by two independent interval-arithmetic checkers, in Arb and Julia, and Lean proves that these ball certificates then exclude four MUBs. The triple classification, meanwhile, is verified by a single certifier in Arb and exact arithmetic. 

Inside each ball, we use the graph-search tests introduced in OpenAI's proof~\cite{OAIMUB2026} to certify where the vectors of a further basis can lie. We reprove the inequalities behind them in Lean, so our proof is logically independent of theirs and relies on none of their certificates~\cite{OAIMUB2026,OAIFourier2026}.

\section{Covering all Hadamard matrices}
\label{sec:nofour}

A set of MUBs can be written as a set of unitary matrices, with each matrix representing one basis and its columns the basis vectors. A complex Hadamard matrix of order $d$ is a $d\times d$ matrix $H$ with entries of unit modulus, $\abs{H_{ij}}=1$, and orthogonal columns, $H^\dagger H = d I_d$~\cite{TadejZyczkowski2006,BengtssonEtAl2007}. Two bases $A$ and $B$ are then mutually unbiased, in the sense of Eq.~\eqref{eq:mub}, exactly when $\sqrt{d}\,A^\dagger B$ is a complex Hadamard matrix. Applying the same unitary to every basis preserves unbiasedness, so without loss of generality we take the first basis to be the identity $I_6$. Every mutually unbiased pair then has the form $\{I_6, H/\sqrt6\}$ and every triple the form
\begin{equation}
    \{I_6,\, H/\sqrt6,\, K/\sqrt6\} \,,
\end{equation}
where $H$, $K$ and $H^\dagger K/\sqrt6$ are complex Hadamard matrices.

A hypothetical quartet of MUBs (which we will prove does not exist) can be written as $\{I_6,H/\sqrt6,K/\sqrt6,L/\sqrt6\}$, where $L$, $H^\dagger L/\sqrt6$ and $K^\dagger L/\sqrt6$ are also Hadamard.
We call the pair $(I_6,H/\sqrt6)$ \emph{extendable} if such $K$ and $L$ exist. If $H$ is extendable, any member of its equivalence class is, and the converse also holds. Two Hadamard matrices are \emph{equivalent}, written $H' \sim H$, if one can be obtained from the other by permuting rows and columns and multiplying them by phases. This means that they lie in the same equivalence class, $H' \in [H]$, where $[H]$ is the set of all Hadamard matrices related to $H$ by such operations. Likewise, two sets of MUBs are equivalent if they are related by a common unitary, by rephasing and permuting the vectors within each basis, and by permuting the bases.

\subsection{Minimal four-phase reconstruction}
\begin{figure}[t]
 \centering
 \includegraphics[width=0.85\columnwidth]{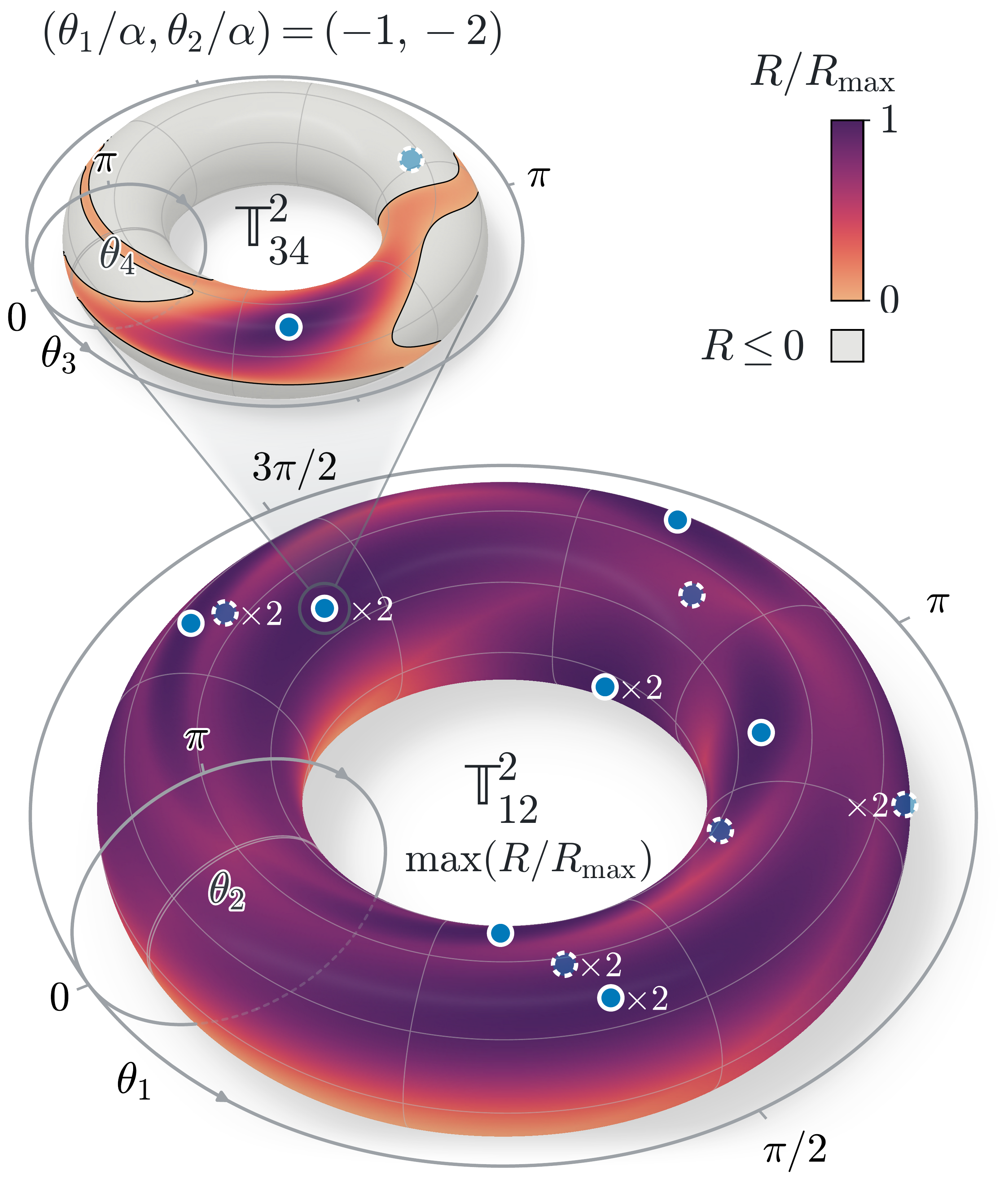}
 \caption{\textbf{Visualization of the reconstruction invariant $R(\theta)$ and its maxima over the four-seed space $\T^4$.} 
 The large torus $\T^2_{12}$ retains the first two seed phases. The smaller torus $\T^2_{34}$ resolves the other two angles at  fixed $(\theta_1/\alpha,\theta_2/\alpha)$ with $\alpha=\arccos(1/4)$. 
 Blue circles mark the 18 seeds which maximize $R(\theta)$ with $R_{\rm max} = 3^{13} / 2^{12}$. In the projection to $\T^2_{12}$, some of them overlap. Dotted outlines correspond to coordinates on the other side of the tori. The large surface shows $M=\max_{\theta_3,\theta_4}(R/R_{\max})$, while the small surfaces shows $R/R_{\max}$, where gray regions correspond to $R \le 0$. Both share the same colorbar.}
 \label{fig:phase}
\end{figure}

We write the complex Hadamard matrix $H$ in block form $H=\left(\begin{smallmatrix}E&B\\C&D\end{smallmatrix}\right)$ with $3\times3$ submatrices. 
We are free to dephase $H$ such that the entries of the first row and column are unity. The top left corner then leaves us with four free phases:
\begin{equation}
E(\theta)=\begin{pmatrix}1&1&1\\1&e^{{\rm i}\theta_1}&e^{{\rm i}\theta_2}\\1&e^{{\rm i}\theta_3}&e^{{\rm i}\theta_4}\end{pmatrix}.
 \label{eq:seed}
\end{equation}
This \emph{seed} $\theta=(\theta_1,\theta_2,\theta_3,\theta_4)$ lies in $\T^4=\T^2_{12}\times\T^2_{34}$, where $\T=\{z\in\C:\abs{z}=1\}$. We recently proved~\cite{wuttig2026completeclassificationcomplexhadamard} that a representative of \textit{every} complex $6\times6$ Hadamard matrix can be built from an appropriate seed and a finite choice of algebraic completions, and that, except for Tao's matrix and a single Karlsson matrix, solving one quadratic and one cubic equation suffices to reconstruct those matrices. 
Here we simplify this four-phase construction further. 

\paragraph*{The reconstruction invariant $R$.}
For a fixed corner $E$, orthogonality requires a block $B$ with entries of magnitude one and $BB^\dagger=6I_3-EE^\dagger$. This, however, does not determine the equivalence class of $B$ uniquely. To distinguish this, we use the complex invariant
\begin{equation}
 \Psi(B)=\frac{\operatorname{perm}(B)\det(B)^2}{\prod_{j,k}B_{jk}},
 \label{eq:phase-invariant}
\end{equation}
where $\operatorname{perm}$ denotes the permanent and $\operatorname{det}$ the determinant. This quantity is unchanged by rephasing or permuting the rows or columns of $B$. From the initial seed $E$, the formulas in the Methods section define three real quantities $g,\eta,\chi$, which satisfy $g=|\det B|^2$, $\eta=|\operatorname{perm}B|^2$ and $\chi=-2\operatorname{Re}\Psi(B)$ for every $B$. Thus $E$ fixes both the real part and the absolute value of $\Psi(B)$, with $|\Psi(B)|^2=g^2\eta$. We now define
\begin{equation}
 R(\theta)=4g^2\eta-\chi^2,
\end{equation}
which is a trigonometric polynomial in the four seed phases.
Permuting the rows or columns of $E$ and dephasing leaves $R$ unchanged.
For a \textit{strict} seed with $R(\theta)>0$, the imaginary part of $\Psi(B)$ has fixed nonzero magnitude but undetermined sign, giving two conjugate values $\Psi_\pm=(-\chi\pm {\rm i}\sqrt{R(\theta)})/2$. Their squared separation is then
\begin{equation}
 R(\theta)=|\Psi_+-\Psi_-|^2
     =4\bigl(\operatorname{Im}\Psi(B)\bigr)^2.
 \label{eq:scalar-meaning}
\end{equation}
For a strict seed $\theta$, the reconstruction in the Methods section realizes both values and shows that each determines $B$ up to column phases and permutations. Choosing the complex conjugate $\Psi(C)=\overline{\Psi(B)}$ fixes $C$ up to row phases and permutations, while orthogonality uniquely forces the last block $D=-CE^\dagger(B^\dagger)^{-1}$.
This conjugate pairing guarantees that $D$ has unit-modulus entries, completing the Hadamard matrix. In the following, we write $H_+(\theta)$ for the block completion with corner $E(\theta)$ obtained by choosing $\Psi(B)=\Psi_-$ and $\Psi(C)=\Psi_+$.

\paragraph*{Which Hadamard matrices are reached by a strict seed completion.} The Fourier matrix $F_n$ is defined through its entries $(F_n)_{jk}=e^{2\pi {\rm i} jk/n}$ for $j,k=0,\ldots,n-1$. Every order-six Hadamard matrix not equivalent to $F_6$ has a strict corner $E$ with $R>0$ (SM Proposition~\ref{prop:critical}), while $F_6$ has $R=0$ at every corner. At a strict corner, both sides $B$ and $C$ carry one of the two distinct, complex-conjugate roots $z_\pm$ of the quadratic $z^2 - \frac{\chi}{g}z + \eta = 0$, so their labels are either conjugate, $\Psi(C)=\overline{\Psi(B)}$ (matched), or equal. Equal labels occur only for Tao's matrix (SM Proposition~\ref{prop:nonmatched}).

Suppose a matrix $M$ has a strict corner with equal labels. We show in SM Proposition~\ref{prop:nonmatched} that $M$ must then contain a $3\times3$ Hadamard subblock. The only order-six Hadamard matrices with such a subblock are, up to equivalence, Tao's matrix and the two Fourier families. Every strict corner of a Fourier-family matrix is matched, so $M$ is equivalent to Tao's matrix.

It remains to fix the orientation of a matched completion. Exchanging both block rows and both block columns replaces the horizontal side $B$ by $C$, which reverses the sign of $\operatorname{Im}\Psi(B)$ and preserves $R>0$. After dephasing, the opposite corner therefore represents the same class with the $H_+$ orientation. This leads us to the following theorem:

\begin{theorem}[Minimal four-phase reconstruction]\label{thm:reconstruction}
For every strict seed $\theta$, the matched completion $H_+(\theta)$ exists and is unique up to equivalence, and its class depends continuously on $\theta$. Every complex Hadamard matrix of order six outside the equivalence classes of Tao's matrix and $F_6$ is equivalent to $H_+(\theta)$ for some strict seed $\theta$.
\end{theorem}
Theorem~\ref{thm:reconstruction} is proved in SM Sec.~\ref{sec:reconstruction-connectedness}. The reconstruction reaches the whole three-dimensional Karlsson family~\cite{Karlsson2011H2,Karlsson2011ThreeParameter} with $F_6$ as a limit~\cite{wuttig2026completeclassificationcomplexhadamard}. Note that $\theta\mapsto[H_+(\theta)]$ is not an injective construction, because different strict seeds can produce equivalent Hadamard matrices.

\subsection{Connectedness of non-Tao Hadamard matrices}
Write $\mathcal X = \mathcal{H}_6 \setminus \{[\mathcal{T}]\}$
for the set of equivalence classes of order-six Hadamard matrices other than Tao's matrix $\mathcal{T}$, where $\mathcal{H}_6$ contains all order-six Hadamard equivalence classes. Two classes are close if they have representatives with nearly equal phases. An explicit distance metric between two equivalence classes is defined in the Methods in Eq.~\ref{eq:methods-phase-metric}.
A space is \emph{connected}, in the topological sense, if it cannot be divided into two nonempty, disjoint open subsets. Equivalently, its only subsets that are both open and closed are the empty set and the set itself.

\begin{theorem}[Connectedness]\label{thm:connected}
The space $\mathcal X$ of non-Tao order-six Hadamard classes is connected.
\end{theorem}

We first prove that the strict seeds form a connected region $\Omega=\{\theta:R(\theta)>0\}$. Consider any connected component of $\Omega$. Its closure in the phase torus $\mathbb{T}^4$ is compact, so $R$ attains a maximum there. This maximum is positive and cannot occur on the boundary, where $R=0$. It therefore occurs inside the component and is a positive local maximum of $R$. Thus every connected component of $\Omega$ contains at least one positive local maximum.

There are exactly eighteen such maxima (SM Proposition~\ref{prop:maxima}), all related by row and column permutations of the corner followed by dephasing. They are shown explicitly in Fig.~\ref{fig:phase}. One such maximal corner is produced by the seed
\begin{equation}
    \theta^* = (\alpha, 2\alpha, -\alpha, -2\alpha)\,,
\end{equation}
where $\alpha=\arccos(1/4)$, giving  $R_{\max}=3^{13}/2^{12}$.
We join these eighteen seeds by explicit paths along which $R$ stays positive. Each path moves the four corner phases from a maximum to one of nine closed curves on which $R=256$. One of these curves is parametrized by
\begin{equation}
 \theta_c(\varphi)=(\varphi,\varphi+\pi,\varphi+\pi,\varphi)
 \pmod{2\pi}\,,
 \label{eq:omega-joining-circle}
\end{equation}
where $0\le\varphi\le2\pi$.
Its nine distinct images under row and column permutations followed by dephasing meet at real corners with entries $\pm1$. These joins connect all nine curves and therefore all eighteen maxima while $R$ remains positive. The path formulas, their strict positivity and the full joining argument are given in SM Sec.~\ref{sec:connectedness}. Since every component contains a maximum and all maxima have been joined, $\Omega$ is connected. The continuous reconstruction $f_+(\theta)=[H_+(\theta)]$ therefore has a connected image $\mathcal A=f_+(\Omega)$.

We can connect the Fourier class (where $R = 0$) to this connected image via the continuous Hadamard family
\begin{equation}
 H(t)=
 \begin{pmatrix}
 F_3&F_3\\
 F_3D(t)&-F_3D(t)
 \end{pmatrix},
\end{equation}
where $D(t)=\diag(1,e^{{\rm i}t},1)$ and $0\le t\le\pi/2$.
For $t>0$, the side-label calculation in SM Sec.~\ref{sec:reconstruction} gives $[H(t)]\in\mathcal A$.
At $t=0$, $H(0)=F_2\otimes F_3\sim F_6$.
This path attaches $[F_6]$ to $\mathcal A$, so $\mathcal A\cup\{[F_6]\}$ is path connected.

Tao's class is isolated and distinct from $[F_6]$~\cite[Lemma~3.3 and Sec.~5.6.4]{TadejZyczkowski2006}, so it cannot belong to the connected set $\mathcal A\cup\{[F_6]\}$ but, by Theorem~\ref{thm:reconstruction}, all other equivalence classes do. Thus we can identify $\mathcal{X} = \mathcal A\cup\{[F_6]\}$, proving Theorem~\ref{thm:connected}.

\subsection{Phase-ball coverage}

\begin{figure*}[t]
 \centering
    \includegraphics[width=0.8\textwidth]{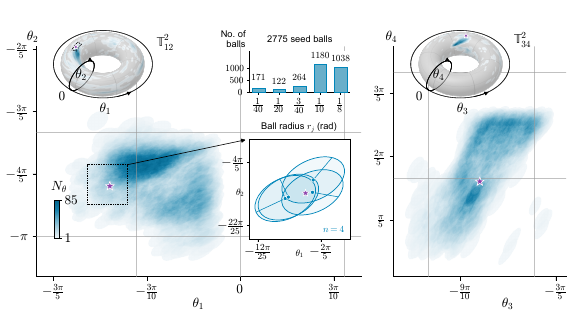}
    \caption{\textbf{Phase-ball coverage in seed space.}
    We fix the maximizing seed $(-\alpha,-2\alpha,-2\alpha,2\pi-4\alpha)$, with $\alpha=\arccos(1/4)$, as the reference (purple stars). The tori show the $2{,}775$ stored four-dimensional seed balls projected onto $(\theta_1,\theta_2)$ and $(\theta_3,\theta_4)$, respectively. The larger plots show coverage near the maximizing seed. Color represents the number of overlapping balls $N_\theta$. Gray (in the tori) and white (in the plots) indicate zero. The lower inset shows all $n=4$ balls in the plotted collection that cover the fixed seed in four dimensions, projected as blue ellipses. Dots mark their centers, and lines extend to their boundaries. The upper inset shows the distribution of phase ball radii $r_j$.}
 \label{fig:atlas}
\end{figure*}

Connectedness now lets us prove global statements about order-six Hadamard matrices from local statements on overlapping regions of Hadamard classes. We build closed regions of classes by reconstructing matrices as their four seed parameters vary, and on each region we certify a local statement: that every triple inside is of a known type, or that no quartet exists.
We then verify that interior parameter points produce open sets of classes, and that every class on the boundary of one region is also represented by an interior point of another. The union $W$ of the regions, with their complex conjugates, transposes and conjugate transposes, is therefore open, and as a finite union of closed sets it is also closed. Since $\mathcal X$ is connected, a nonempty $W$ must equal $\mathcal X$, so the statements certified on each region hold throughout $\mathcal X$.

\paragraph*{Seed balls.}
For each region $j$, we choose a central seed $\theta^{(j)}\in\Omega$, a vector of four corner angles. We vary these angles by adding $M_jw$, where $w\in\mathbb R^4$ contains four real displacement parameters and the fixed invertible $4\times4$ matrix $M_j$ sets the directions and scales of these changes. Allowing $w$ to have Euclidean length at most $r_j>0$ gives the closed ellipsoid
\begin{equation}
 B_j=\{\theta^{(j)}+M_jw:\|w\|_2\le r_j\} \subset \Omega,
 \label{eq:seed-ball}
\end{equation}
which we call a \emph{seed ball}, with $w=0$ giving its center. Since $\Omega$ is open, every sufficiently small seed ball lies inside it. We choose $r_j$ accordingly and certify $B_j\subset\Omega$ by showing that $\operatorname{Im}\Psi(B)$ keeps one sign throughout the closed ball, so that $R=4(\operatorname{Im}\Psi)^2>0$ there. The same certificate shows that the matched completion exists throughout the ball with a fixed orientation. Using interval arithmetic, we then bound the reconstructed matrices over the whole closed ball, including its boundary, so that a single local test applies to the entire family (Methods). Different corners can represent the same Hadamard class, so covering $\mathcal X$ does not require every equivalent seed to lie in a stored ball.

Figure~\ref{fig:atlas} illustrates the arrangement of the certified seed balls through projections onto the two pairs of corner phases, with enlarged views around a maximum of $R$. The gaps in these projections need not leave Hadamard classes uncovered, because an equivalent corner can represent the same class inside a certified ball. Most of the $2{,}775$ stored balls have parameter radii $\frac{1}{10}$ or $\frac{1}{8}$ radians. 

\paragraph*{The neighbourhood of $F_6$.}
The Fourier matrix needs its own phase ball. Its corners all have $R=0$, so no strict seed ball contains it. We therefore add a chart through $F_6$ whose coordinates are three shifted corner phases and $y=-2\operatorname{Im}\Psi(B)$, so that $y^2=R$. These coordinates continue across $R=0$ and place $F_6$ at the origin. We certify one closed ball in this chart and classify every triple that extends a pair $\{I_6,H/\sqrt6\}$ with $H$ in this image.

\paragraph*{Boundary transfers.}
To show that the union $W$ is open, we need two facts. First, the interior of each ball maps to an open set of classes. Nearby Hadamard matrices keep the matching signs of a strict corner, so by matched uniqueness an open set of strict seeds maps to an open set of classes. The chart at $F_6$ has the same property across $R=0$ (Methods). Second, every class on the boundary of a ball must lie in the interior of another. We cover each parameter sphere $\|w\|_2=r_j$ by spherical caps, $117{,}391$ in all, whose union is checked exactly, and on each cap certify a transfer into the interior of another ball or of the $F_6$ chart. A transfer either keeps the same corner, or changes corner and checks that the matching signs agree, or enters the chart at $F_6$, with row and column permutations, transposition and conjugation allowed throughout.

\paragraph*{Building the seed ball cover.}
To construct the seed ball cover, we build an automated driver program. Starting from an initial ball, it repeatedly takes boundary boxes that no existing ball covers yet. For each one it proposes a new seed center, ellipsoid and transfer target, reconstructs the corresponding Hadamard matrices and runs the local tests. If a test fails, the driver reduces the radius $r_j$ and tries again. Separate certifying programs then re-check every enclosure, local test and boundary transfer from the stored data alone~\cite{Tindall2026NoFourMUBsRepository}.

\section{No four mutually unbiased bases}
\subsection{Classifying all triples}

The known MUB triples form a continuous two-parameter family and a single isolated triple built from the Fourier matrix~\cite{Goyeneche2013,MatszangoszSzollosi2024}. We now prove that these exhaust all triples up to equivalence, as conjectured by Matolcsi, Matszangosz, Varga and Weiner~\cite{MatolcsiEtAl2026MUBTriplets}.

Tao's matrix admits no third basis~\cite{BrierleyWeigert2009}, so triples must live in $\mathcal X$. For all seed balls and around $F_6$, we certify that every triple whose second basis $H/\sqrt6$ lies in it either does not exist or has an order-three symmetry (Methods). In four balls the certificates cover only the labelling of the bases in which $H$ is the transition closest to Fourier form, as measured by a score defined in Methods. Every triple has such a labelling, so none is missed. The certificates use rigorous enclosures of the vectors unbiased to both $I_6$ and $H/\sqrt6$, valid for every $H$ in the region at once, and subdivide a region where needed (see Methods and SM Sec.~\ref{sec:tc-classification}).

The symmetry is a unitary $T$ with $T^3=I_6$ that fixes each vector of one basis up to a phase and permutes the vectors of each of the other two in two cycles of three. Taking the fixed basis to be $I_6$, it is diagonal,
\begin{equation}
 T=\operatorname{diag}(1,1,\omega,\omega,\omega^2,\omega^2),\qquad \omega=e^{2\pi {\rm i}/3}.
 \label{eq:triple-symmetry}
\end{equation}
Because each eigenvalue appears twice, any vector $v$ unbiased to $I_6$ satisfies $\langle v,Tv\rangle=0$, so $\{v,Tv,T^2v\}$ is automatically orthonormal. Each of the other two bases is therefore fixed by just two vectors, and the triple by twelve phases. Solving the remaining unbiasedness condition shows that every triple with this symmetry, independently of the phase balls, is in the two-parameter family or is the isolated Fourier triple (Methods and SM Sec.~\ref{sec:tc-recognition}). Since we have certified that all of our phase balls (and the chart around $F_6$) force every triple to have the symmetry, this proves Conjecture~1 of Ref.~\cite{MatolcsiEtAl2026MUBTriplets}.

Both types of triple contain a pair related by a matrix in the Fourier family, which has been excluded from a quartet~\cite{Jaming2009}. This gives our first proof of $N(6)=3$. We therefore, in combination with Theorem~\ref{thm:reconstruction}, can prove the following theorem.

\begin{theorem}[Classification of all MUB sets]\label{thm:all-mubs}
Up to equivalence, every nonempty set of MUBs in $\mathbb C^6$ is one of the following:
\begin{enumerate}
\item a single basis $\{I_6\}$;
\item a pair $\{I_6,H/\sqrt6\}$, where $H$ is Tao's matrix, $F_6$, or $H_+(\theta)$ for a strict seed $\theta$;
\item a triple in the continuous two-parameter family, or the isolated Fourier triple.
\end{enumerate}
No set contains four or more bases.
\end{theorem}

\subsection{Exclusion of quartets}
The same phase ball cover can also be used to exclude quartets directly without classifying triples. In each seed ball, we ask whether $I_6$ and $H/\sqrt6$ can be joined by two further bases. Every vector of such a basis is unbiased to $I_6$, so it is fixed by five phases~\cite{Grassl2004,BrierleyWeigert2009,FuhrRzeszotnik2015,AnderssonBengtsson2017}. Interval arithmetic encloses all vectors that are also unbiased to $H/\sqrt6$, for every $H$ in the ball at once, in finitely many small boxes. We group these boxes into \emph{labels}, small enough that no label contains two orthogonal or two unbiased vectors, and build a graph whose edges record which pairs of labels could still contain orthogonal or unbiased vectors. Two further bases would then need twelve distinct labels, split into two groups of six, with every pair within a group joined by an orthogonality edge and every pair across the groups joined by an unbiasedness edge. An exhaustive search finds no such pattern in any ball (SM Proposition~\ref{prop:finite}). Around $F_6$, a separate exclusion rules out quartets for every Hadamard matrix within a phase ball (Methods). Since Tao's matrix admits no third basis~\cite{BrierleyWeigert2009}, this proves $N(6)=3$.

This local test follows OpenAI's recent approach~\cite{OAIMUB2026}, although our connectedness result (Theorem~\ref{thm:connected}) allows us to replace their nine-dimensional cover of $249{,}878$ centres with $2{,}775$ balls in four seed phases.

\section{Outlook and Discussion}
Every order-six complex Hadamard matrix other than $F_6$ and Tao's matrix can be built from four phases, and the space of these matrices is topologically connected outside of Tao's matrix. 
Connectedness lets us turn questions about the whole space to checks on a finite cover of certified four-dimensional phase balls. We exploit this to obtain a complete classification of all sets of MUBs in dimension six. 

First, every MUB pair $\{I_6, H / \sqrt{6}\}$ is either formed from Tao's matrix or belongs to this connected space which can be navigated by continuously varying just four phases. Second, every triple $\{I_6, H / \sqrt{6}, K / \sqrt{6}\}$ is equivalent to a member of a continuous two-parameter family or to the isolated Fourier triple, proving the conjecture of Matolcsi, Matszangosz, Varga and Weiner~\cite{MatolcsiEtAl2026MUBTriplets}. Neither type of triple admits a fourth basis~\cite{Jaming2009}, which proves Zauner's conjecture $N(6)=3$~\cite{Zauner1991,Zauner1999}. The same cover also excludes quartets directly, giving a second proof that no four MUBs exist, without using the triple classification. Dimension six is thus the first non-prime-power dimension in which all sets of MUBs are classified.

The order-six Hadamard reconstruction is useful beyond classifying MUBs. Its identities give an analytical proof of the Fourier-vanishing conjecture~\cite{MatolcsiRuzsaWeiner2013} for which OpenAI previously reported a computer-assisted proof~\cite{OAIFourier2026}, see SM Proposition~\ref{prop:fourier-vanishing}. More broadly, the four-phase coordinates and the connectedness of $\mathcal X$ give a compact four-dimensional parameter space, covered by finitely many phase balls, from which to prove statements about order-six Hadamard matrices.

\subsection{Mutually unbiased constellations} 
Every triple of MUBs in dimension six is strongly unextendible: no unit vector is unbiased to all three bases. This directly follows from the classification here, since both types of triple are known to be strongly unextendible~\cite{Sun2026Completion,McNultyWeigert2012Triples}. In the notation of mutually unbiased constellations~\cite{BrierleyWeigert2008}, where $\{n_1,\ldots,n_4\}_6$ denotes four groups of $n_i$ orthonormal rays with every squared overlap between groups equal to $1/6$, this means that $\{5,5,5,1\}_6$ is empty. Since five orthonormal rays fix the sixth, a five is a complete basis. This resolves Problem~10.3 of Ref.~\cite{McNultyWeigert2026} (SM Sec.~\ref{sec:strong-unextendibility}).

Smaller constellations do exist, however. We give exact algebraic examples of $\{5,3,3,3\}_6$, $\{5,4,3,1\}_6$ and $\{5,5,2,2\}_6$ (SM Sec.~\ref{sec:constellation-certificates}). The first had not been found by numerical searches~\cite{BrierleyWeigert2008} and was believed to not exist~\cite{BrierleyWeigert2010SDP,WeigertDurt2010,Mortimer2023}.

\subsection{Higher dimensions}
After six, the smallest open dimensions are now ten and twelve. Our bound does not pass to multiples of six, because restricting bases to a six-dimensional subspace need not preserve orthonormality or unbiasedness, and dimension twelve already supports five MUBs, more than the tensor-product count of four~\cite{wuttig2026mutuallyunbiasedbases}. Dimension ten, where the tensor product of dimensions two and five gives three MUBs, is the natural next target for the strategy adopted here: a corner reconstruction of the Hadamard matrices, a scalar whose maxima can be used to prove the space connected, and certified local exclusions from overlapping regions. Proving all, or almost all, order ten Hadamard matrices obey a strict seed completion from a dephased $5 \times 5$ corner would be a first step towards such a proof. 
More generally, establishing an upper bound or specific value for non-prime power dimensions $N(d) < d+1$ (and thus $N(d) < d$~\cite{Weiner2013}) would be a major step towards a complete understanding of MUBs~\cite{McNultyWeigert2026}.

\subsection{Physical implications}
Our results describe how much freedom complementary measurements in dimension six have, and that freedom shrinks as more measurements are added. A pair of complementary measurements is flexible: since $\mathcal X$ is connected (Theorem~\ref{thm:connected}), any mutually unbiased pair not involving Tao's matrix can be deformed continuously into any other, up to equivalence, without losing mutual unbiasedness. Specifically, one basis can be held fixed while the four-phase reconstruction gives explicit coordinates for the other along the path. Tao's matrix is the only exception: it is an isolated point that no such deformation reaches.

Adding a third measurement removes most of this freedom. Pairs that extend to a triple lie in two-parameter families inside the four-parameter space of pairs, so a generic pair admits no third. Every triple is equivalent to a member of one two-parameter family or to a single isolated triple, so protocols built on three mutually unbiased measurements, such as three-basis key distribution or entropic uncertainty relations, can be optimised over an explicit, low-dimensional set. Even a single extra state is too much: strong unextendibility means that no pure state gives uniform outcome statistics in all three measurements of any triple.

A fourth measurement is therefore impossible. No four orthonormal-basis measurements in dimension six can all be mutually unbiased, even under ideal experimental control, which rules out saturating quantum random access code and Bell-correlation benchmarks that require four MUBs~\cite{AguilarEtAl2018QRAC,PratColomer2022}. A system with more distinguishable states can thus support fewer mutually complementary measurements, as $N(5)=6$, $N(6)=3$ and $N(7)=8$ show, and dimension six now proves that the tensor-product lower bound can be exact~\cite{Klappenecker2004,wuttig2026mutuallyunbiasedbases}.

\section{Methods}
\label{sec:methods}
\subsection{Four-phase reconstruction}
\label{sec:methods-reconstruction}
We write an order-six complex Hadamard matrix in terms of $3\times3$ blocks as $H=\left(\begin{smallmatrix}E&B\\C&D\end{smallmatrix}\right)$ and dephase it, so that its first row and first column consist of ones. The top-left corner is then fixed by four phases $\theta=(\theta_1,\theta_2,\theta_3,\theta_4)$,
\begin{equation}
 E(\theta)=\begin{pmatrix}
 1&1&1\\
 1&e^{{\rm i}\theta_1}&e^{{\rm i}\theta_2}\\
 1&e^{{\rm i}\theta_3}&e^{{\rm i}\theta_4}
 \end{pmatrix},
\end{equation}
and the first row of $B$ and the first column of $C$ also consist of ones. Orthogonality of the first three rows and of the first three columns of $H$ fixes the overlap matrices of $B$ and $C^T$ in terms of $E$~\cite{Szollosi2012FourParameter,wuttig2026completeclassificationcomplexhadamard},
\begin{equation}
\begin{aligned}
 G_h &= BB^\dagger = 6I_3-EE^\dagger,
 \\ G_v &= C^T\bar C = 6I_3-E^T\bar E,
\end{aligned}
\label{eq:methods-grams}
\end{equation}
with $\bar E$ the complex conjugate of $E$. The diagonal entries of $G_h$ all equal $3$, and its off-diagonal entries $s=(G_h)_{21}$, $t=(G_h)_{31}$ and $r=(G_h)_{32}$ are
\begin{equation}
 \begin{aligned}
 s&=-(1+e^{{\rm i}\theta_1}+e^{{\rm i}\theta_2}),\\
 t&=-(1+e^{{\rm i}\theta_3}+e^{{\rm i}\theta_4}),\\
 r&=-(1+e^{{\rm i}(\theta_3-\theta_1)}+e^{{\rm i}(\theta_4-\theta_2)}).
 \end{aligned}
 \label{eq:methods-reconstruction-overlaps}
\end{equation}
The reconstruction invariant $R$ of the main text depends only on $X=|s|^2$, $Y=|t|^2$, $Z=|r|^2$ and the combinations
\begin{equation}
 \begin{aligned}
 p&=X+Y+Z,\\
 J&=2\operatorname{Re}(sr\bar t).
 \end{aligned}
 \label{eq:methods-reconstruction-combinations}
\end{equation}
Explicitly,
\begin{equation}
 \begin{aligned}
 R(E(\theta))&=4g^2\eta-\chi^2\\
 &=4(27-3p+J)^2(9-p+J)\\
 &\quad-\bigl[XY+XZ+YZ-XYZ- \\
 &\quad\quad\quad (p-9)(p-18)+(p-15)J\bigr]^2,
 \end{aligned}
 \label{eq:methods-reconstruction-invariant}
\end{equation}
with $g=27-3p+J$, $\eta=9-p+J$, and $\chi$ the expression in square brackets. For a block $B$ with $BB^\dagger=G_h$ and unit-modulus entries, $g=|\det B|^2$ and $\eta=|\operatorname{perm}B|^2$. Both overlap matrices give the same $g$, $\eta$, $\chi$ and $R$, and for $R>0$ both are positive definite with $g>0$ (Supplementary Proposition~\ref{prop:sides}). Equation~\eqref{eq:methods-reconstruction-invariant} therefore evaluates $R$ directly from the four seed phases.

\paragraph*{The side blocks.}
We write
\begin{equation}
 B=\begin{pmatrix}
 1&1&1\\
 x_1&x_2&x_3\\
 z_1&z_2&z_3
 \end{pmatrix},
 \qquad |x_j|=|z_j|=1,
\end{equation}
so that $BB^\dagger=G_h$, which requires
\begin{equation}
 \sum_{j=1}^3x_j=s,\qquad
 \sum_{j=1}^3z_j=t,\qquad
 \sum_{j=1}^3z_j\bar x_j=r.
 \label{eq:methods-reconstruction-row-sums}
\end{equation}
For $R>0$ these equations have two solutions up to column permutations, distinguished by the invariant $\Psi$ of the main text (see Eq.~\ref{eq:phase-invariant}), which takes one of the values $\Psi_\pm=(-\chi\pm {\rm i}\sqrt R)/2$. The completion $H_+(\theta)$ uses the conjugate pair
\begin{equation}
 \Psi(B)=\Psi_-,\qquad \Psi(C)=\Psi_+.
\end{equation}
To construct $B$ from this choice, set $\zeta=-\Psi(B)/g=(\chi+{\rm i}\sqrt R)/(2g)$. Since $g>0$, this is the root with positive imaginary part of
\begin{equation}
 \zeta^2-\frac{\chi}{g}\zeta+\eta=0,
\end{equation}
whose roots satisfy $\zeta+\bar\zeta=\chi/g$ and $|\zeta|^2=\eta$. From the overlap matrix alone we form
\begin{equation}
\begin{aligned}
 P_1&=rs^2+\bar r\,t^2+(3-Z)st,\\
 P_2&=t\bar s^2+r^2\bar t+(3-Y)r\bar s,\\
 P_3&=s\bar t^2+\bar r^2\bar s+(3-X)\bar r\,\bar t,
\end{aligned}
\end{equation}
and, with $(c_1,c_2,c_3)=(9-Z,9-Y,9-X)$, the unit-modulus numbers
\begin{equation}
 \lambda_j=\frac{g+c_j\bar\zeta}{\overline{P_j}},
 \qquad j=1,2,3,
\end{equation}
whose denominators are nonzero for $R>0$ (Supplementary Proposition~\ref{prop:sides}). These give the row products $u=x_1x_2x_3$ and $v=z_1z_2z_3$ and the permanent $\operatorname{perm}B$,
\begin{equation}
 u=\lambda_1\bar\lambda_2,\qquad
 v=\lambda_1\bar\lambda_3,\qquad
 \operatorname{perm}B=\zeta\lambda_1,
\end{equation}
consistent with $|\operatorname{perm}B|^2=|\zeta|^2=\eta$.

Each row of $B$ now follows from its sum and product. Since the entries have unit modulus, $x_jx_k=u\bar x_\ell$ for distinct $j,k,\ell$, so $\sum_{j<k}x_jx_k=u\bar s$, and similarly $\sum_{j<k}z_jz_k=v\bar t$. The entries of the second and third rows are therefore the roots of
\begin{equation}
 \begin{aligned}
 \xi^3-s\xi^2+u\bar s\,\xi-u&=0,\\
 \xi^3-t\xi^2+v\bar t\,\xi-v&=0,
 \end{aligned}
\end{equation}
respectively, counted with multiplicity. To assemble the columns, we fix the order of the $x_j$ and choose the order of the $z_j$ satisfying
\begin{equation}
 \sum_{j=1}^3x_jz_j=st-\operatorname{perm}B,\qquad
 \sum_{j=1}^3\bar x_jz_j=r,
\end{equation}
where the first condition follows from $\left(\sum_jx_j\right)\left(\sum_kz_k\right)=\sum_jx_jz_j+\operatorname{perm}B$. This determines $B$ uniquely up to column permutations, including when roots repeat (Supplementary Proposition~\ref{prop:sides}). The same procedure applied to $G_v$ with $\bar\zeta$ gives $C^T$, and hence $C$ up to row permutations.

\paragraph*{The forced block.}
Since $g=|\det B|^2>0$, $B$ is invertible, and orthogonality between the first and last three rows of $H$ determines
\begin{equation}
 D=-CE^\dagger G_h^{-1}B.
 \label{eq:methods-forced}
\end{equation}
For the conjugate pair of labels, $D$ has unit-modulus entries and thus $H$ is a complex Hadamard matrix (Supplementary Proposition~\ref{prop:matched}). Reversing both labels gives the other matched completion.

\subsection{Bounds on reconstructed matrices}
\label{sec:methods-phase-balls}
We work in $\mathcal X$, the space of equivalence classes of order-six complex Hadamard matrices, with Tao's isolated class removed. Here equivalence allows row and column rephasings and permutations. Each class $[H]$ represents a pair of MUBs via $\{I_6, H / \sqrt{6}\}$.

We measure the distance between two classes through the phases of their entries. Writing $H=e^{{\rm i}Q_H}$ entrywise, with a real phase matrix $Q_H\in\mathbb R^{6\times6}$, we define
\begin{equation}
 d_{\rm ph}([H],[K])=\inf\,\|Q_H-Q_K\|_F,
 \label{eq:methods-phase-metric}
\end{equation}
where the infimum runs over all representatives $e^{{\rm i}Q_H}\sim H$ and $e^{{\rm i}Q_K}\sim K$, that is, over row and column rephasings and permutations and over the choice of each phase modulo $2\pi$, and $\|\cdot\|_F$ denotes the Frobenius norm.

We choose a central seed $\theta_0 \in \mathbb{T}^4$ and describe nearby seeds using four real parameters $w=(w_1,\ldots,w_4)$:
\begin{equation}
 \theta(w)=\theta_0+M_{\theta_0} w,\qquad \|w\|_2\le r_H,
 \label{eq:methods-phase-constraints}
\end{equation}
Here $r_H$ is the chosen radius of the ball. The point $w=0$ gives the central seed, and the invertible $4 \times 4$ matrix $M_{\theta_0}$ maps $w$ to an ellipsoid of seed angles centered around $\theta_0$.

For each $w$ in the ball, we reconstruct the order-six Hadamard matrix for the seed $\theta(w)$ by computing $B,C$ as above and $D$ from Eq.~\eqref{eq:methods-forced}, choosing $\Psi(C)=\overline{\Psi(B)}$ with $\operatorname{Im}\Psi(B)>0$. This is the other matched completion, $\overline{H_+(-\theta(w))}$: the stored balls use this orientation throughout, and since the cover includes conjugates (and, by the block exchange of SM Sec.~\ref{sec:non-tao-space}, either orientation reaches every class), nothing below depends on the choice. Let $Q(w)$ be its phase matrix, with row and column phases removed by the projection in Supplementary Information, section~\ref{sec:phase-distance}. Using interval arithmetic, we verify that $\operatorname{Im}\Psi(B)$ keeps one sign throughout the closed ellipsoid. This gives $R=4(\operatorname{Im}\Psi(B))^2>0$ there, and the same matched completion at every seed, so that one enclosure
\begin{equation}
 \begin{gathered}
 Q(w)=P+Uw+\nu(w),\qquad \|w\|_2\le r_H,\\
 \nu(w)\in\mathbb R^{6\times6},\qquad \|\nu(w)\|_F\le n,
 \end{gathered}
 \label{eq:methods-certified-phase-enclosure}
\end{equation}
can describe every Hadamard matrix reconstructed from the ellipsoid.
Here $P$ is a stored finite precision approximation of the center phases $Q(0)$, and $U:\mathbb R^4\to\mathbb R^{6\times6}$ a finite precision approximation of their linear change with $w$. The remainder $\nu(w)=Q(w)-P-Uw$ collects the nonlinear variation and the differences between these approximations and the exact values, and $n$ bounds its Frobenius norm over the whole ball. The radius $r_H$ thus bounds the four seed parameters, while the enclosure bounds all $36$ entry phases. The ellipsoid, the reconstruction choice and the enclosure bounds are stored in the certificates we provide~\cite{Tindall2026NoFourMUBsRepository}.

To obtain $n$, we start from a quadratic approximation of the completion and solve independent combinations of the row-orthogonality equations for a correction that leaves the seed fixed. Bounds on the quadratic term, the correction and rounding then give $n$ (Supplementary Information, section~\ref{sec:phase-ball-proof}).

\subsection{Fourier neighborhood}
\label{sec:methods-fourier-chart}
The pair $\{I_6,F_6/\sqrt6\}$ is already known not to belong to a quartet~\cite{Grassl2004,Jaming2009}. No strict seed ball contains $F_6$, because every corner of $F_6$ has $R=0$, so we add a separate parameterized patch through $F_6$, whose image lies within phase distance $0.1636$ of $F_6$. We classify every triple on this patch, and exclude quartets (by an exact, Lean-formalized cover) on the larger ball of phase radius $7/40$ around $F_6$, which contains it. The reconstruction, exclusion and openness bounds are given in Supplementary Information, section~\ref{sec:fourier-chart-proof}.

\subsection{Representatives of all MUB triples and the order-three symmetry}
\label{sec:methods-triples}
Up to equivalence, a triple of MUBs can be written as $\{I_6,U_2,U_3\}$, with $H=\sqrt6\,U_2$ and $K=\sqrt6\,U_3$ in the notation of the main text. Its three transition matrices, $H$, $K$ and $H^\dagger K/\sqrt6$, are all Hadamard. Below, $\widehat F_n=F_n/\sqrt n$ is the normalized Fourier matrix and $\omega=e^{2\pi {\rm i}/3}$.

The isolated Fourier triple is the tensor product of qubit and qutrit MUB triples,
\begin{equation}
 \begin{aligned}
 U_2&=\widehat F_2\otimes\widehat F_3,\\
 U_3&=[\operatorname{diag}(1,{\rm i})\,\widehat F_2]\otimes
       [\operatorname{diag}(1,\omega^2,\omega^2)\,\widehat F_3].
 \end{aligned}
 \label{eq:methods-isolated-triple-bases}
\end{equation}
The continuous two-parameter family is~\cite{Szollosi2010Hypocycloids,Jaming2009}
\begin{equation}
 \begin{aligned}
 U_2&=\frac1{\sqrt2}
 \begin{pmatrix}\widehat F_3&D\widehat F_3\\\widehat F_3&-D\widehat F_3\end{pmatrix},\\[2pt]
 U_3&=\frac1{\sqrt2}
 \begin{pmatrix}
 R\widehat F_3&-DS^\dagger\widehat F_3\\
 S\widehat F_3&DR^\dagger\widehat F_3
 \end{pmatrix},
 \end{aligned}
 \label{eq:methods-explicit-continuous-bases}
\end{equation}
where the diagonal matrices $D,R,S$ are given explicitly as functions of two seed coordinates $(\theta_1,\theta_2)$ in SM Sec.~\ref{app:triple-seeds}.

Both of these types of triple have an order-three symmetry: a unitary $T$ with $T^3=I_6$ that maps each of the three bases to itself. It keeps every vector of one basis fixed up to a phase, and cycles the vectors of each of the other two bases in two groups of three. Up to equivalence, we may take the basis that $T$ keeps fixed to be $I_6$. A unitary that fixes every standard basis vector up to a phase is diagonal, and here its diagonal entries are the three cube roots of unity, each appearing twice. Ordering the coordinates accordingly,
\begin{equation}
 T=\operatorname{diag}(1,1,\omega,\omega,\omega^2,\omega^2).
\end{equation}

Let $\{I_6,H/\sqrt6,K/\sqrt6\}$ be a triple on which $T$ acts in this way. Label rows by $(t,a)$, where $t\in\mathbb Z_3$ picks an eigenspace of $T$ and $a\in\mathbb Z_2$ one of its two coordinates, and columns by $(s,b)$, where $b\in\mathbb Z_2$ picks a three-cycle and $s\in\mathbb Z_3$ a position in it. After row and column rephasings,
\begin{equation}
\begin{aligned}
H_{ta,sb}&=\omega^{ts}(-1)^{ab}x_t^b,\\
K_{ta,sb}&=\omega^{ts}(-1)^{ab}y_t^b
\begin{cases}
u_t,&a=0,\\
v_t,&a=1,
\end{cases}
\end{aligned}
\label{eq:triple-cyclic}
\end{equation}
with twelve unit phases $x_t,y_t,u_t,v_t$. These formulas already make $H$ and $K$ Hadamard and the columns of $H^\dagger K/\sqrt6$ orthogonal, so it remains to impose unit modulus on the entries of $H^\dagger K/\sqrt6$. This matrix consists of four circulant $3\times3$ blocks, and a discrete Fourier transform reduces the conditions to equations in their Fourier sums. Generic solutions give the continuous family of Eq.~\eqref{eq:methods-explicit-continuous-bases}. The exceptional solutions have upper circulant blocks with first rows $\omega^{\pm s^2+ls}$, $l\in\mathbb Z_3$, up to constant phases, and give either the continuous family again or the isolated triple of Eq.~\eqref{eq:methods-isolated-triple-bases}. All degenerate parameter values are included (SM Sec.~\ref{sec:tc-recognition}).

\subsection{Local triple classification and coverage}
\label{sec:methods-local-exclusion}
We treat one parameter ball of the cover at a time (Sec.~\ref{sec:methods-phase-balls}). By Sec.~\ref{sec:methods-triples}, it suffices to show that every triple $\{I_6,H/\sqrt6,K/\sqrt6\}$ with $H$ reconstructed from the ball either does not exist or has the symmetry $T$ (in four balls, every such triple whose first transition has the least $\Gamma_F$; see \emph{Ordering} below). Where the tests below are inconclusive on the whole ball, we subdivide its enclosing box in $w$ (see Eq.~\eqref{eq:methods-phase-constraints}) into four-dimensional \emph{parameter cells} and repeat them on each cell. For about four-fifths of the seed balls, the tests exclude every completing basis $K$, so $\{I_6,H/\sqrt6\}$, with $H$ reconstructed from the ball, cannot extend to a triple.

\paragraph*{1. Find every vector unbiased to $I_6$ and $H/\sqrt6$.}
Each column of $K/\sqrt6$ is a vector with entries $e^{{\rm i}\phi_j}/\sqrt6$, with $\phi_1=0$. Writing $H_{jk}=e^{{\rm i}Q_{jk}}$, unbiasedness to $H/\sqrt6$ requires
\begin{equation}
f_k(\phi,Q)=\frac16\left|\sum_{j=1}^{6}e^{{\rm i}(\phi_j-Q_{jk})}\right|^2-1=0,\qquad k=1,\ldots,6.
\label{eq:methods-vector-equations}
\end{equation}
We cover the five phases $\phi_2,\ldots,\phi_6$ by closed \emph{vector phase boxes}, and discard a box when verified bounds, uniform over every $H$ in the ball or cell (Sec.~\ref{sec:methods-phase-balls}), show that these equations have no solution in it. The retained boxes are grouped into \emph{labels}, each small enough to contain no two orthogonal vectors, so the columns of a completing basis lie in six distinct labels. We call such a choice of six labels a \emph{pattern}, and keep every pattern that passes the pairwise orthogonality tests on all fifteen pairs. These tests follow OpenAI's approach~\cite{OAIMUB2026} (SM Sec.~\ref{sec:vector-exclusion}).

\paragraph*{2. Exclude or classify each pattern.}
Pairwise tests do not show that six columns can be chosen consistently. Write $F=0$ for the full set of real orthogonality and unbiasedness equations of the triple. Throughout, $H$ ranges over the ball or cell, and all six columns are tested against the same $H$. Each pattern is resolved by one of the following.
\begin{itemize}
\item \emph{Exclusion.} The pattern's boxes are subdivided and the pairwise tests repeated. A box is removed when it is incompatible with every box of some other label, and the pattern is excluded when no six compatible boxes remain. Alternatively, a verified bound $y^TF>0$ for a fixed real vector $y$ excludes it (SM Sec.~\ref{sec:tc-joint}).
\item \emph{Normal-form localisation.} Near the continuous family, two directions of the column phases barely affect the equations to first order. About a numerical reference close to a triple, we separate these two directions from the other $28$, and remove the leading quadratic and cubic influence of these two directions on the $28$ by an invertible change of variables, bounding what remains. Verified bounds on the remaining terms, which include the variation of $H$ and do not assume that the reference is exact, confine every solution to a small region. Boxes outside it are discarded, and the pattern goes to the next test with these tighter bounds (SM Sec.~\ref{sec:tc-weak-normal-form}).
\item \emph{Symmetry.} We propose an anchor basis and a column ordering for which the symmetry, conjugated into that basis, should cycle the vectors of the other two bases in two groups of three. The proposed action preserves $F$, and verified derivative bounds show that the phase distance $\varepsilon$ of any solution from these cyclic relations satisfies $\varepsilon\le q\varepsilon$ with $q<1$. Hence $\varepsilon=0$, and every triple in the pattern has the symmetry (SM Eq.~\eqref{eq:tc-symmetry}). This does not assert that a triple exists.
\item \emph{Ordering.} In four of the $2{,}775$ seed balls, patterns along a nearly flat direction are not resolved by the tests above at the cell sizes used. There a pattern may instead be discharged by the score $\Gamma_F$ of SM Eq.~\eqref{eq:gamma-score}, which is invariant under equivalence, transposition and conjugation. Specifically, if one of the other two transitions has a score strictly below $\Gamma_F(H)$ throughout the pattern, the pattern contains no triple whose first transition has the least score of the three (SM Lemma~\ref{lem:minimum-transition}).
\end{itemize}

\paragraph*{3. Verify coverage.}
A parameter cell is discarded only when it is proved to lie strictly outside the original ball, and each retained cell is identified with the ball's reconstruction branch (SM Sec.~\ref{sec:tc-subdivision}). On every retained cell, every pattern is excluded, shown to be symmetric or, in the four balls above, discharged by ordering. To see that no triple is missed, order its bases so that the transition between the first two has the least $\Gamma_F$ of the three. This transition is not Tao's matrix, since that matrix admits no third basis~\cite{BrierleyWeigert2009}, so it has a representative in the cover, where the steps above classify every triple with this ordering. Together with the classification of pairs and the exclusion of quartets, this gives Theorem~\ref{thm:all-mubs}.

\section*{Acknowledgements}
M.C.W. acknowledges the hospitality of the Center for Computational Quantum Physics at the Flatiron Institute, a division of the Simons Foundation.

\paragraph*{AI assistance and author contributions.}
OpenAI's ChatGPT 6 Astra and Claude Opus 5.5 assisted significantly in developing the proof. M.C.W. and J.T. directed the AI agents in a search for a cover-based proof excluding four MUBs based on the four-dimensional coordinates of Hadamard matrices. The authors reviewed the mathematical arguments and methods developed by the agentic systems, and refined them with a view towards simplicity and efficiency. The authors prepared the manuscript and are responsible for its content.

\section*{Data availability and verification}
All code and data are available at \url{https://github.com/JoeyT1994/NoFourMUBS}. The repository contains the Lean formalisation and the certifiers, and follows the order of this manuscript: from the four-phase reconstruction and connectedness theorems to the phase-ball cover, the exclusion of quartets and the classification of triples. The certificates for the triple classification, including those for the ball at $F_6$, are attached to the release \href{https://github.com/JoeyT1994/NoFourMUBS/releases/tag/tc-verification-2026-11-04}{\texttt{tc-verification-2026-11-04}}. SM Table~\ref{tab:verification} summarizes how each result is verified.

\paragraph*{Phase-ball cover and quartet exclusion.}
Two independent checkers, one using Arb ball arithmetic~\cite{Johansson2017Arb} and one written in Julia, verify from the stored data that the $2{,}775$ seed balls and the ball at $F_6$ cover every non-Tao class: the enclosure of each ball, and the $117{,}391$ boundary caps through which every ball's boundary transfers into another. The same checkers also verify the exclusion of quartets on every seed ball. At $F_6$, they check that the chart lies within phase distance $7/40$ of $F_6$, where quartets are excluded by an exact, Lean-formalized cover. The complete check took about $33$ minutes on a $128$-core node in Arb, and $67$ minutes on a $64$-core node in Julia.

\paragraph*{Triple classification.}
On the same phase-ball cover, each of the $2{,}775$ seed balls has its own certificate for its triples, $3.3$~GB in total, and the ball at $F_6$ is classified by a separate calculation whose $383{,}829$ shell verdicts are also release assets. The certifier reruns the tests described in the Methods on each ball using only the data stored in its certificate, recomputing every bound in Arb or exact arithmetic, apart from a few floating-point comparisons in the graph-search tests that carry proved error margins (SM Sec.~\ref{sec:tc-verification}). For $2{,}373$ of the phase balls, the tests close the whole ball at once, and checking takes about $20$~s on one core. The other $402$ need refining into smaller cells, so their certificates are dyadic trees, with a median size of $2.5$~MB and a median check of about $90$~s on one $128$-core node. Checking all $2{,}775$ took about $900$ core-hours, under five hours on sixteen nodes of the Flatiron Institute's Rusty supercomputer.
\bibliography{ref}

\clearpage
\onecolumngrid
\appendix*
\setcounter{figure}{0}
\renewcommand{\thefigure}{A\arabic{figure}}
\renewcommand{\theHfigure}{appendix.\arabic{figure}}

\let\NoFourMUBsSupplementEnd\relax
\expandafter\ifx\csname @classoptionslist\endcsname\relax
  \def\NoFourMUBsSupplementEnd{\end{document}}
  \documentclass[a4paper,10pt]{article}
  \usepackage[margin=20mm]{geometry}
  \usepackage[T1]{fontenc}
  \usepackage{amsmath,amssymb,amsthm,booktabs}
  \usepackage[hidelinks]{hyperref}
  \newtheorem{proposition}{Proposition}
  \renewcommand{\theproposition}{S.\arabic{proposition}}
  \renewcommand{\theequation}{S.\arabic{equation}}
  \renewcommand{\thesubsection}{\arabic{subsection}}
  \providecommand{\Real}{\operatorname{Re}}
  \providecommand{\Imag}{\operatorname{Im}}
  \providecommand{\perm}{\operatorname{perm}}
  \providecommand{\diag}{\operatorname{diag}}
  \providecommand{\tr}{\operatorname{tr}}
  \begin{document}
  \raggedbottom
  \section*{Supplementary Information}
\fi

\providecommand{\paperref}[1]{\ref{#1}}
\providecommand{\papereqref}[1]{\eqref{#1}}
\newtheorem{supplemma}{Lemma}
\renewcommand{\thesupplemma}{S.\arabic{supplemma}}
\providecommand{\restated}[2]{\par\medskip\noindent\textbf{#1}\par\nopagebreak\noindent\textit{#2}\par\medskip}

Table~\ref{tab:verification} summarizes how each of the major results in this work is verified.

\begin{table}[h]
\centering
\small
\begin{tabular}{@{}p{0.29\linewidth}p{0.44\linewidth}p{0.23\linewidth}@{}}
\toprule
Result & How it is verified & File location \\
\midrule
Reconstruction and connectedness (Theorems~\paperref{thm:reconstruction} and~\paperref{thm:connected}) & Exact algebra and certificates, fully formalized in Lean & \nolinkurl{MinimalReconstruction/}, \nolinkurl{Connectedness/}, \nolinkurl{lean/} \\
The seed balls, the Fourier chart and their symmetry images cover every non-Tao class & Interval arithmetic, checked by two independent programs in Arb and Julia & \nolinkurl{PhaseCover/} \\
No four MUBs exist & The above Arb and Julia programmes exclude quartets in every seed ball. Meanwhile, an exact cover formalized in Lean excludes them near $F_6$, and Lean then derives $N(6)=3$ under the atlas hypotheses checked by Arb and Julia & \nolinkurl{QuartetExclusion/}, \nolinkurl{lean/} \\
Triples on $2{,}771$ seed balls and minimum-transition triples on the other four are of known type & One certificate per ball, checked in Arb and exact arithmetic, with binary64 bounds for Stage~C & \nolinkurl{TripleClassification/} \\
Every triple near $F_6$ is of known type & A certified core, the two crossings and a shell of cells, checked in Arb and exact arithmetic, with the same Stage~C bounds & \nolinkurl{TripleClassification/} \\
Fourier vanishing (Proposition~\ref{prop:fourier-vanishing}) & Exact identities of the reconstruction, formalized in Lean & \nolinkurl{MinimalReconstruction/}, \nolinkurl{lean/} \\
\bottomrule
\end{tabular}
\caption{How each of the major results in this work is verified and where the corresponding files can be found on the corresponding \href{https://github.com/JoeyT1994/NoFourMUBS/tree/tc-verification-2026-11-04}{Github repository}.}
\label{tab:verification}
\end{table}

\renewcommand{\contentsname}{Supplemental contents}

\subsection{Proofs of the reconstruction and connectedness theorems}
\label{sec:reconstruction-connectedness}
This section proves the two global results of the main text. We first find the positive local maxima of $R$ on the seed torus and join them by explicit paths of strict seeds, so that the strict seeds form a connected set (Sec.~\ref{sec:connectedness}). We then construct the two side blocks and the forced fourth block at every strict seed, and show that the completion is unique and continuous (Sec.~\ref{sec:reconstruction}). Section~\ref{sec:exceptional-classes} shows that the only classes not reached in this way are those of $F_6$ and Tao's matrix, and Sec.~\ref{sec:proof-theorems-1-2} combines these results.

\restated{Theorem~\paperref{thm:reconstruction} (Minimal four-phase reconstruction; restated)}{For every strict seed $\theta$, the matched completion $H_+(\theta)$ exists and is unique up to equivalence, and its class depends continuously on $\theta$. Every complex Hadamard matrix of order six outside the equivalence classes of Tao's matrix and $F_6$ is equivalent to $H_+(\theta)$ for some strict seed $\theta$.}

\restated{Theorem~\paperref{thm:connected} (Connectedness; restated)}{The space $\mathcal X$ of non-Tao order-six Hadamard classes is connected.}

\subsubsection{Positive maxima and connected strict seeds}
\label{sec:connectedness}
\phantomsection\label{proof:omega-connectedness}
Every component of $\Omega=\{\theta\in\mathbb T^4:R(\theta)>0\}$ contains a positive local maximum, and we join all maxima.

Use two ordered triples of unit phases, each with product one:
\begin{equation}
 E(\mathbf e,\mathbf o)=
 \begin{pmatrix}1&1&1\\1&e_1/o_2&o_1/e_3\\1&o_1/e_2&e_1/o_3\end{pmatrix}.
 \label{eq:methods-products-to-seed}
\end{equation}
For $(a,b,c,d)=(e^{{\rm i}\theta_1},e^{{\rm i}\theta_2},e^{{\rm i}\theta_3},e^{{\rm i}\theta_4})$, take $\mathbf e=(ad,b,c)/\tau$,
$\mathbf o=(bc,d,a)/\tau$, $\tau^3=abcd$.
The traces $U=\sum e_k$, $V=\sum o_k$ range independently over
\begin{equation}
 \Delta(z)=|z|^4+18|z|^2-8\Real(z^3)-27\le0,\qquad
 \partial\{\Delta\le0\}=\{2e^{{\rm i}t}+e^{-2{\rm i}t}\}.
 \label{eq:deltoid}
\end{equation}
The cubic $\lambda^3-U\lambda^2+\bar U\lambda-1$ recovers the first triple up to ordering, and similarly for $V$. Roots vary continuously with the trace, including coincidences, and the map from triple phases to seed phases has determinant three. Thus a seed maximum is also a maximum in $U,V$.

\begin{proposition}[Positive local maxima]\label{prop:maxima}
The positive local maxima of $R$ are the eighteen seeds obtained by row and column permutations and dephasing of
\begin{equation}
E_*=\begin{pmatrix}1&1&1\\1&e^{{\rm i}\alpha}&e^{2{\rm i}\alpha}\\1&e^{-{\rm i}\alpha}&e^{-2{\rm i}\alpha}\end{pmatrix},
 \qquad \alpha=\arccos(1/4),\qquad R(E_*)=\frac{3^{13}}{2^{12}}.
 \label{eq:maximum}
\end{equation}
They all attain the global maximum.
\end{proposition}

\begin{proof}
We consider interior and boundary traces in $\Delta\le0$; this boundary differs from $R=0$. Put $A=U+V$, $B=U-V$, $u=|A|^2$, $v=|B|^2$. Then
\begin{equation}
 g=2v-3u,\qquad \eta=v-2u,\qquad
 \chi=\tfrac12(3u-v)(v-u+12)-2\Real(AB^2).
 \label{eq:trace-invariants}
\end{equation}

If $\Delta(U)<0$ and $\Delta(V)<0$, both traces can vary freely near the point, and $R>0$ gives $v>2u$.
If $\chi=0$, $\partial_vR=4(2v-3u)(6v-11u)>0$.
If $A=0$, rotating $B$ to $\sqrt v$ and varying $A=a$ gives
$\partial_aR=-2v^2(v+12)<0$.
Otherwise $\partial_\phi R=-4\chi v\sqrt u\sin\phi$, where $\phi=\arg(AB^2)$, so $\phi=0$ or $\pi$ at a positive critical point.
The $R$-preserving rotation $(A,B)\mapsto(e^{-2{\rm i}\tau}A,e^{{\rm i}\tau}B)$ gives real $A=a\ne0$ and $B=\sqrt v$, with
\begin{equation}
 f(a,v)=4(2v-3a^2)^2(v-2a^2)
 -\left[\tfrac12(3a^2-v)(v-a^2+12)-2va\right]^2.
 \label{eq:maxima-interior-polynomial}
\end{equation}
Eliminating $v$ from $f_a=f_v=0$ leaves, apart from $R=0$,
\begin{equation}
 9a^6-135a^5+854a^4-2838a^3+5070a^2-4392a+1296=0.
 \label{eq:maxima-interior-eliminant}
\end{equation}

Starting with this polynomial and its derivative, take the negative remainders of successive polynomial divisions. The number of sign changes drops by two from $-\infty$ to $+\infty$. Since the polynomial is positive for $a\le0$, both real roots are positive. Substitution into $f_a=f_v=0$ gives $\chi<0$ at both. The variation $A=ae^{{\rm i}\varepsilon}$ has $R''=-4\chi av>0$, so neither is a maximum.

Suppose $\Delta(U)=0$ at a smooth boundary point while $\Delta(V)<0$.
The same rotation preserves $R$ and gives $U_\tau=(e^{-2{\rm i}\tau}A+e^{{\rm i}\tau}B)/2$.
If the derivative of $\Delta(U_\tau)$ at zero is nonzero, one sign of $\tau$ moves $U_\tau$ inside, reducing to the preceding case.
If that derivative vanishes, write $U=2z+z^{-2}$ and $V=rz-(r+1)z^{-2}$, with $|z|=1$ and $r\in\mathbb R$. Then
$\left.\frac{d^2}{d\tau^2}\Delta(U_\tau)\right|_0
=-9(r+2)^2(2-z^3-z^{-3})^2/2<0$,
so small nonzero $\tau$ again moves $U_\tau$ inside.
Equality requires three coincident phases or the other trace on the boundary.
For three coincident phases, take $U=3$, $V=x+{\rm i}y$.
At positive $R$, $\partial_yR=0$ forces $y=0$ and $-1<x<-4/5$, where $\partial_xR<0$.

With both traces on the boundary, set
$c=\cos((t-s)/2)$, $h=\cos(3(t+s)/2)$ for
$U=2e^{{\rm i}t}+e^{-2{\rm i}t}$, $V=2e^{{\rm i}s}+e^{-2{\rm i}s}$. Then

\begin{equation}
 R=-256(2c^2+1)^2(8c^4+4c^3h-4c^2+1)^2(16c^4+16c^3h+4c^2h^2-4c^2-4ch-1).
 \label{eq:boundary-maximum-polynomial}
\end{equation}
At $|c|=1$, $R\le0$; at $c=0$ the only critical point is the saddle $(0,0)$.
For $c\ne0$, elimination in the interior gives
$(4c^2-1)(2c^2-1)(16c^4-4c^2-1)=0$.
The first two factors give $R=0$. The positive case of the last has
$c^2=(1+\sqrt5)/8$, $h=4c^3-3c$.
At its representative $V=3$, $\cos t=(\sqrt5-3)/4$, splitting the repeated phases by
$U(\varepsilon)=2e^{{\rm i}t}\cos\varepsilon+e^{-2{\rm i}t}$ gives
$R''(0)=330(23\sqrt5-51)>0$.
On $h=1$, positivity requires $-1/2<c<1/2$; the same polynomial root count for $R_c(c,1)$ leaves only $c=1/4$.
The case $h=-1,c=-1/4$ gives the same traces. Ordering the triples gives the eighteen seeds below. A global maximum is positive and must be among them, so permutation invariance proves the proposition.
\end{proof}

\paragraph*{Joining the maxima.}
To join the maxima through $R>0$, let $\mathbf e^{(i)}(t)$ have entry $e^{-2{\rm i}t}$ at position $i$ and $e^{{\rm i}t}$ elsewhere, and let $\theta_{ij}(t,s)$ be the seed of
$E(\mathbf e^{(i)}(t),\mathbf e^{(j)}(s))$.
The $3\times3$ positions and two signs of $\alpha$ give the eighteen distinct maxima:
\begin{equation}
 \theta_{ij}(\alpha,-\alpha),\quad\theta_{ij}(-\alpha,\alpha),
 \qquad i,j=1,2,3.
 \label{eq:methods-maximum-path-endpoints}
\end{equation}
Join each along $\theta_{ij}(t,-t)$ to $t=\pm\pi/2$. For $c=\cos t\in[0,1/4]$,
\begin{equation}
 R(\theta_{ij}(t,-t))
 =256(1-2c)(1+2c)^5(1+2c^2)^2(4c^3-2c+1)^2>0.
 \label{eq:methods-omega-path-positivity}
\end{equation}
The endpoints lie on the nine curves $\{\theta_{ij}(t,t+\pi)\}$, where $R=256$.
For $i=j=1$ the entries are $(z,-z,-z,z)$, $z=-e^{-3{\rm i}t}$, giving the main-text curve.
At $(t,s)=(0,\pi)$ all curves with fixed $j$ meet because $\mathbf e=(1,1,1)$; at $(\pi,0)$ all with fixed $i$ meet because $\mathbf o=(1,1,1)$. Thus all nine curves and eighteen maxima are joined. Hence $\Omega$ is connected and, being open in four phase coordinates, path connected.

\subsubsection{Reconstruction of the side blocks}
\label{sec:reconstruction}
We write $H[I,J]$ for the submatrix with row set $I$ and column set $J$, both in increasing order, where $I^c,J^c$ denote their complements in $\{1,\ldots,6\}$.

\phantomsection
\label{sec:seed-invariants}
Use the overlaps and invariants of Methods Eqs.~\papereqref{eq:methods-reconstruction-overlaps}--\papereqref{eq:methods-reconstruction-invariant}, writing $z=-\Psi(S)/g$ for the reconstruction value $\zeta$. At a strict seed,
\begin{equation}
 z_\pm=\frac{\chi\pm {\rm i}\sqrt R}{2g}=-\frac{\Psi_\mp}{g}.
 \label{eq:methods-label-roots}
\end{equation}

For the main-text family $H(t)=\left(\begin{smallmatrix}F_3&F_3\\F_3D(t)&-F_3D(t)\end{smallmatrix}\right)$, with $D(t)=\diag(1,e^{{\rm i}t},1)$ and $0\le t\le\pi/2$, take $I=\{2,4,6\}$ and $J=\{1,3,5\}$. Then
\begin{equation}
 \Psi(H(t)[I,J^c])=-3e^{-2{\rm i}t}(2e^{{\rm i}t}-1)^3
 =\overline{\Psi(H(t)[I^c,J])}.
 \label{eq:methods-fourier-labels}
\end{equation}
Its imaginary part is $-6\sin t(1+\cos t)<0$ for $t>0$. Hence $R=144\sin^2t(1+\cos t)^2>0$, giving a strict corner with both blocks and opposite signs. This supplies the starting point for existence below and the path to $F_6$ at $t=0$.

\label{sec:side-constructor}
For either overlap matrix $G$ in Methods Eq.~\papereqref{eq:methods-grams}, use its $s,t,r,X,Y,Z$ and the reconstruction formulas in Methods. Here $P_j,c_j,\lambda_j$ for $j=1,2,3$ denote the corresponding Methods quantities with $\zeta=z$, where $z$ is a root from Eq.~\eqref{eq:methods-label-roots}.
For $R>0$ the denominators are nonzero and $|\lambda_j|=1$, as shown below. A unit-entry block with $BB^\dagger=G$ satisfies
\begin{equation}
 g=|\det B|^2,\qquad \eta=|\perm B|^2,\qquad
 \chi=-2\Real\Psi(B),\qquad |\Psi(B)|^2=g^2\eta.
 \label{eq:methods-side-identities}
\end{equation}

\begin{proposition}[Reconstruction of $B$ and $C$]\label{prop:sides}
For $R(\theta)>0$, both overlap matrices (Methods) are positive definite, $g>0$, and each root determines a block with the required overlap matrix and entries of magnitude one, unique up to column phases and permutations. Construct $C$ by applying this to $C^T$. Equations~\eqref{eq:methods-side-identities} and~\papereqref{eq:scalar-meaning} also hold at $g=0$; only $z$ requires $g>0$.
\end{proposition}
\begin{proof}
Expanding after multiplying by products of the unit entries proves the identities without assuming $g>0$.
For $R>0$, $\eta+|\det E|^2=36-4p$ and $g=\eta+2(9-p)$ give $p<9$, $g>0$.
The principal minors of the overlap matrix $3,9-X,g$ are positive. Also,
\begin{equation}
 |P_j|^2=g^2+c_j\chi+c_j^2\eta
 =\eta\left(c_j+\frac{\chi}{2\eta}\right)^2+\frac{R}{4\eta}>0,
\end{equation}
which gives $|\lambda_j|=1$ using $z+\bar z=\chi/g$ and $|z|^2=\eta$. 
To prove existence, write a block with first row $(1,1,1)$ as $S(\phi)$, where $\phi$ gives the six remaining phases. Its overlap map and, by differentiation, its Jacobian are
\begin{equation}
\begin{aligned}
 \gamma(\phi)&=(\Real s,\Imag s,\Real t,\Imag t,\Real r,\Imag r),\\
 |\det D\gamma(\phi)|&=\tfrac14|\Imag\Psi(S(\phi))|,
\end{aligned}
 \label{eq:side-jacobian}
\end{equation}
Fix $\Psi(B)=\Psi_-$, $\Psi(C)=\Psi_+$. The nonzero Jacobians make the seeds admitting both blocks open in $\Omega$. Compactness of the normalized block phases makes this set closed: limits satisfy the overlap equations and retain the chosen roots, whose imaginary parts cannot vanish in $\Omega$. It is nonempty by Eq.~\eqref{eq:methods-fourier-labels}, so connectedness gives existence throughout $\Omega$. Conjugating the construction at the conjugate seed gives the other roots. The preceding connectedness proof used only $R$.

To prove uniqueness, set the first row to one and write the block as $S=(\mathbf1;x;y)$. Put $u=\prod_jx_j$, $v=\prod_jy_j$. The same identities give
\begin{equation}
 u=\lambda_1\bar\lambda_2,\qquad
 v=\lambda_1\bar\lambda_3,\qquad \perm S=z\lambda_1.
 \label{eq:side-products}
\end{equation}
The cubics in Methods, with $y_j$ in place of $z_j$, fix the entries of each row up to ordering.
Choose an ordering of $x$. The quantities $\sum_jy_j=t$, $\sum_jx_jy_j=st-\perm S$ and $\sum_j\bar x_jy_j=r$ are fixed. If the $x_j$ are distinct, the coefficient matrix of these three equations for $y$ is invertible: multiplying column $j$ by $x_j$ gives a Vandermonde matrix. If $x=(a,a,b)$ with $a\ne b$, the first two sums fix $y_3$ and $y_1+y_2$, while $v$ fixes $y_1y_2$. Only an interchange of the first two columns remains. All three $x_j$ cannot coincide because $g>0$. This proves uniqueness up to column permutations, including when entries coincide.
\end{proof}

\begin{proposition}[Completion and continuity]\label{prop:matched}
For any reconstructed blocks $B,C$ with roots $(z_+,z_-)$ or $(z_-,z_+)$, the formula for the fourth block $D$ gives entries of magnitude one and completes the strict corner to a Hadamard matrix. The first choice defines a continuous map $f_+$ from $\Omega$ to Hadamard equivalence classes.
\end{proposition}

\begin{proof}
Block orthogonality gives $HH^\dagger=6I_6$.
For normalized $B,V=C^T$ with roots $z,w$, put
$W=\bar V(|D_{kl}|^2-1)_{kl}\bar B^T$.
Each row and column of $D$ has squared norm three, so the first row and column of $W$ vanish.
The sums $\sum_kS_{ak}\overline{S_{ck}S_{ik}}$ for a normalized block $S$ are affine in its root by Eq.~\eqref{eq:side-products} and $z^2=(\chi/g)z-\eta$.
Substituting these expressions into $W$ and reducing $z^2,w^2$ with their quadratic equations expresses each remaining entry as a linear combination of $z+w-\chi/g$ and $zw-\eta$, with coefficients depending only on the seed.
For conjugate roots,
\begin{equation}
 z+w=\chi/g,\qquad zw=\eta
 \quad\Longrightarrow\quad W=0.
 \label{eq:completion-defect-factorization}
\end{equation}

Since $B,V$ are invertible, $W=0$ forces every entry of $D$ to have magnitude one.
\end{proof}

\phantomsection\label{sec:methods-connectedness}
\paragraph*{Continuity and orientation.}

The nonzero Jacobians in Eq.~\eqref{eq:side-jacobian} give unique analytic local blocks $B,C$ by the implicit function theorem; $g>0$ makes $D$ analytic too. Uniqueness therefore makes $f_+$ continuous in the phase distance of Eq.~\eqref{eq:methods-phase-distance}, with the local analyticity required below.
Corner permutations extend to the full matrix and preserve the roots. Thus all eighteen maximal seeds (see Fig.~\paperref{fig:phase}) give the same Hadamard class.

\subsubsection{Fourier families and exceptional classes}
\label{sec:exceptional-classes}
We exclude equal roots at strict Fourier-family corners and identify their only class with $R=0$ everywhere.
For $F=F_3$ and $v=(1,p,q)$, write~\cite[Eq.~(1)]{MatszangoszSzollosi2024}
$\mathcal F(p,q)=\left(\begin{smallmatrix}F&D_vF\\F&-D_vF\end{smallmatrix}\right)$,
$D_v=\diag(v)$.
Its paired rows $(i,i+3)$ have the same left half and opposite right halves. For every three-element row and column set,
\begin{equation}
 \Psi(\mathcal F(p,q)[I,J^c])
 =\overline{\Psi(\mathcal F(p,q)[I^c,J])}.
 \label{eq:family-matching}
\end{equation}
If the selected rows contain a complete pair, so do the complementary rows. Direct expansion gives $\Psi=0$ when all columns lie in one half and $24$ when both halves are used.
Otherwise choose one member of each pair, absorbing its sign into $v_i$.
If all columns lie in one half, both blocks are equivalent to $F$, with $\Psi=81$.

If both halves are used, column ordering and rephasing give, with $\omega=e^{2\pi {\rm i}/3}$,
\begin{equation}
 E=\begin{pmatrix}1&1&v_1\\1&\omega&v_2\\1&\omega^2&v_3\end{pmatrix},
 \qquad D=\begin{pmatrix}1&-v_1&-v_1\\\omega^2&-\omega v_2&-\omega^2v_2\\\omega&-\omega^2v_3&-\omega v_3\end{pmatrix}.
\end{equation}
Reversing columns makes $D$ equivalent to $\bar E$, proving Eq.~\eqref{eq:family-matching}.

For $p=e^{{\rm i}\alpha}$, $q=e^{{\rm i}\beta}$, expansion and the four sign choices between paired rows give
\begin{equation}
 \max_{\rm corners}R
 =36\bigl(|\sin(2\beta-\alpha)|+|\sin(\alpha+\beta)|
                         +|\sin(\beta-2\alpha)|\bigr)^2.
 \label{eq:fourier-family-max-R}
\end{equation}
It vanishes exactly at $\alpha=a\pi/3$, $\beta=b\pi/3$,
$0\le a,b<6$, $a+b\equiv0\pmod3$.
Pair exchanges change the signs of $p,q$; cyclic right-column permutations multiply them by $(\omega^k,\omega^{2k})$.
These set $p=1$, then $q=\pm1$, whose sign can be removed.
Thus every such matrix is equivalent to $F_2\otimes F_3\sim F_6$.
Transposition gives the same results for the transposed family.

\begin{proposition}[Equal roots at a strict corner]\label{prop:nonmatched}
If $B,C$ at a strict corner do not have conjugate roots, the Hadamard matrix is equivalent to Tao's matrix.
\end{proposition}
\begin{proof}
The two distinct roots are conjugate, so $B,C$ have equal roots. Compare $H$ with the conjugate-root completion $H_0$ having the same first three rows $T=[E\ B]$.

Normalized blocks with different roots cannot share a column. Removing a shared column would leave rank-two matrices with the same overlap matrix, related by a unitary fixing $(1,1)$. It preserves each row's mean and multiplies its difference by a phase $\lambda$. A row $(u,v)$ keeps entries of magnitude one only if
$\Imag\lambda\,\Imag(u\bar v)=0$.
A nonreal row ratio forces $\lambda=\pm1$, giving the same block up to a column interchange and hence the same root. Otherwise all rows have $u=\pm v$, allowing infinitely many blocks, contrary to Proposition~\ref{prop:sides}.

Put $Q=T^\dagger T/6$, $P=I-Q$.
These rank-three projections have diagonal $1/2$. The normalized bottom rows of $H,H_0$, after taking adjoints, give two orthonormal bases of the range of $P$ sharing no ray.
Their two-dimensional spaces of traceless diagonal Hermitian operators intersect only at zero: a common nonzero operator would have a one-dimensional eigenspace. On the range of $P$, the Hermitian operators form a nine-dimensional real space. Each
$Pe_je_j^\dagger P$ is orthogonal to both diagonal spaces under the trace inner product, since its expectation in every basis vector is $1/6$.
Six operators in a space of dimension at most five are dependent, so
$PD_wP=0$ for a nonzero real diagonal $D_w$.

Since $|P_{jk}|=|Q_{jk}|$, the Frobenius norms of $PD_wP,QD_wQ$ agree. Both vanish, giving
\begin{equation}
 QD_w+D_wQ=D_w,\qquad
 (w_j+w_k)Q_{jk}=0\ (j\ne k),\qquad \sum_jw_j=0.
 \label{eq:projection-relation}
\end{equation}
If all $|w_j|$ agree, there are three of each sign; columns of $T$ with the same sign are orthogonal.
Otherwise group indices by $|w_j|$. Each diagonal block of the projection $Q$ has integer trace equal to half its size, so one block has size two. The corresponding columns of $T$ are proportional and orthogonal to the others. Rephase their direction to $(1,1,1)^T$. The other columns are then multiples of the two nonconstant columns of $F_3$, both of which occur because $T$ has rank three. In either case $H$ contains a $3\times3$ Hadamard block.

The classification of order-six Hadamards containing a $3\times3$ Hadamard subblock is related to Ref.~\cite[Lemma~11]{LiangChenLongQiu2021}. We give the argument here with both transpose orientations retained. Normalize this block to $F=F_3$, with $\omega=e^{2\pi {\rm i}/3}$. The other three blocks are also order-three Hadamards. Equivalence to $F$, absorbing cyclic permutations and reversals into phases, gives the following form with unit parameters $v=(1,p,q)$, $w=(1,r,s)$:
\begin{equation}
 H(v,w)=\begin{pmatrix}F&D_vF\\FD_w&-\tfrac13FD_wF^\dagger D_vF\end{pmatrix},
 \quad D_v=\diag(1,p,q),\quad D_w=\diag(1,r,s).
 \label{eq:fourier-block}
\end{equation}
The bottom-right block has squared row and column norms three. Flatness leaves four real conditions whose Fourier sums give
\begin{equation}
 a_\pm(p,q)a_\pm(r,s)=0,\qquad
 a_\pm(p,q)=1/p+\omega^{\pm1}p/q+\omega^{\mp1}q.
 \label{eq:three-point-flatness}
\end{equation}
Both $a_\pm(p,q)$ vanish exactly when $p^3=1$, $q=p^2$, making $v$ a column of $F$. Such a $v$ or $w$ gives a Fourier family or its transpose.
Otherwise each pair makes one function vanish. Its three unit summands have product one, so all parameters are cube roots of unity. Cyclic permutations let us set $p=r=1$, leaving $q,s\in\{\omega,\omega^2\}$ with $q\ne s$. The two choices are transposes. For $q=\omega^2,s=\omega$, row order $(1,4,5,2,6,3)$ gives Tao's symmetric representative.
Equation~\eqref{eq:family-matching} excludes the Fourier families from the assumed equal-root case, leaving Tao's matrix.
\end{proof}

\paragraph*{Singular minors and conjugate rows.}
A singular $3\times3$ matrix with entries of magnitude one has proportional rows or columns. At rank two, writing one row as $\alpha$ times a second plus $\beta$ times a third gives
$2\Real(\bar\alpha\beta z_j)=1-|\alpha|^2-|\beta|^2$, $|z_j|=1$.
If $\alpha\beta\ne0$, a line meets the unit circle at most twice, so two coordinate ratios agree.
In an order-six Hadamard, proportional rows in a minor give complete rows
$(1,1,1;1,1,1)$ and $(1,1,1;-1,-1,-1)$ after rephasing.
The other rows sum to zero on both halves, which are therefore multiples of nonconstant Fourier rows. Orthogonality pairs equal Fourier directions with opposite right phases, giving $\mathcal F(p,q)$; proportional columns give its transpose.

Three complete rows $(\mathbf1;x;\bar x)$ also force a Fourier family.
Orthogonality gives $\sum x_j=\sum x_j^2=0$. Let $e_k$ be the sum of all products of $k$ distinct entries of $x$. Orthogonality and $e_{6-j}=e_6\bar e_j$ give $e_1=e_2=e_4=e_5=0$. Thus the polynomial with the six entries as roots is

\begin{equation}
 \prod_{j=1}^6(T-x_j)=T^6-e_3T^3+e_6.
 \label{eq:conjugate-row-polynomial}
\end{equation}
This is quadratic in $T^3$, so the entries are the three cube roots of each of two numbers. Ordering them puts these rows in $[F\ D_vF]$ with $v=(1,u,u^{-1})$.
Since $a_\pm(u,u^{-1})=\omega^{\pm1}(u^3-1)/u$, Eq.~\eqref{eq:three-point-flatness} makes either $v$ or $w$ a column of $F$, giving a Fourier family or its transpose.

\begin{proposition}[Vanishing of $R$ at every corner]\label{prop:critical}
An order-six Hadamard has $R=0$ at all $400$ three-row, three-column corners if and only if it is equivalent to $F_6$.
\end{proposition}
\begin{proof}
Every corner has $R\ge0$ by Eq.~\papereqref{eq:scalar-meaning}.
If a minor is singular or three rows have the conjugate form above, Eq.~\eqref{eq:fourier-family-max-R} gives $F_6$.
Assume that all corners have $R=0$ and neither preceding reduction applies. We constrain the overlap invariants $p,J$ of a three-row block, then rule out the remaining values using orthogonality of the complete rows.

Normalize three complete rows to $(\mathbf1;x;y)$ and let $E_A$ be the block in column triple $A$. Its overlap entries are
$s=\sum_Ax_j$, $t=\sum_Ay_j$, $r=\sum_A\bar x_jy_j$.
The complementary block has overlaps $-s,-t,-r$, preserving $p$ and reversing $J$.
Put $W=XY+XZ+YZ-XYZ$ and
$D_C=\Psi(E_C)-\overline{\Psi(E_C)}=0$.
For an invertible block with row products $u,v$, define
$\lambda_1=-uv\,\overline{\det E_A}/\det E_A$,
$\lambda_2=\lambda_1/u$, $\lambda_3=\lambda_1/v$.
The identities used for reconstruction give
$(g+c_jz)\lambda_j=P_j$ for the real root $z$.
This equation remains valid when its coefficient vanishes.

If $J=0$, the two equations $R=0$ give
$g=3z^2>0$, $p=9-z^2$, $W=z^2(z+3)^2$, $-3\le z\le3$.
The values $z=-3,-2$ give $p=0,5$.
Otherwise put $\tau=z+3$.
The denominators $z(3\tau-X)$ and their relabellings are nonzero: $X=3\tau$ would give
$(Y-Z)^2=3z(z+2)^2\tau^2/(3z+8)<0$.
Here nonnegativity forces $-3<z<0$ and $3z+8>0$. Relabel and rephase so $s,t>0$.
Order the columns so that $A$ consists of columns $1,2,3$. Write $D_{ij}$ for $D_C$ when column $i$ of this block is replaced by column $j$, where $i=1,2,3$ and $j=4,5,6$.
Expanding the two vanishing sums
\begin{equation}
 \sum_{i=1}^3\sum_{j=4}^6x_jD_{ij}=0,\qquad
 \sum_{i=1}^3\sum_{j=4}^6(s-x_i)y_jD_{ij}=0
 \label{eq:zero-J-mixed-sums}
\end{equation}
with these identities gives
$XYZ=(1-z)\tau^3$. In the separate square-sum elimination over all twenty corners, the moment denominator $(\prod_i(3z+9-X_i))^3$ is nonzero by the case just excluded, and the discarded factor $z^2\tau^4+(z+2)^2XYZ$ is positive.
Consequently
\begin{equation}
 (T-X)(T-Y)(T-Z)=(T-\tau)^2(T-(1-z)\tau).
 \label{eq:zero-J-modulus-factorization}
\end{equation}
Relabel so $X=Y=\tau$ and rephase so $s=t=\sqrt\tau$.
Then $\lambda_1=1$, $\perm E_A=z$ and
$\sum_Ax_jy_j=st-\perm E_A=3$.
Each summand is one, so $y_j=\bar x_j$.
The complementary half has $s'=t'=-\sqrt\tau$ and the same $z$, giving the same conclusion there. This is the conjugate row form already excluded.

For $J\ne0$, exchange halves so $J>0$.
The equations $R(p,\pm J,W)=0$ give
\begin{equation}
 W=\frac{2J^2+p(p-3)(p-9)}{p-15},\qquad
 4J^2+(p-9)^3(p-5)=0.
 \label{eq:nonzero-J-W}
\end{equation}
Since $p+J\le9$, put $k=2J/(9-p)^2>0$.
Then $p=9-4/(1+k^2)$, $J=8k/(1+k^2)^2$.
The row-product denominators on the two halves are
\begin{equation}
 L_{j,\pm}=3(9-p)\pm J-\frac{2(1\pm k)c_j}{1+k^2}.
 \label{eq:nonzero-J-denominators}
\end{equation}
Expanding $\sum_CD_C=0$ gives
$(sr\bar t-\bar s\bar r t)(X-Y)(Y-Z)(Z-X)=0$.
Thus two moduli agree or $sr\bar t$ is real.
If two moduli agree, relabel so $X=Y$. The equation for $W$ gives their common value
$2/(1+k^2)$ or $(3k^2\pm2k+1)/(1+k^2)$.
If $sr\bar t$ is real, $XYZ=16k^2/(1+k^2)^4$.
Substitution in the second vanishing sum
$\sum_C(\sum_{j\in C}x_jy_j)D_C=0$
rules out all four cases when every $L_{j,\pm}\ne0$.
The resulting nonzero factors are
$k^2(3k^4+2k^2+3)$, $k^2(3k^2+1)^2$ and $k^2(1+k^2Z)$.
Hence some denominator vanishes. Factoring their product in these four cases leaves the positive possibilities
\begin{equation}
 k=1/3,\quad k=1,\quad
 3k^3+k^2+k-1=0.
 \label{eq:nonzero-J-denominator-products}
\end{equation}
These give $p=27/5$, $p=7$ and $p=p_*$, respectively; eliminating $k$ in the last case gives the cubic below.
Exchanging halves supplies both signs of $J$.
The possibilities obtained in the two cases are
\begin{equation}
 (p,J)=(0,0),\ (5,0),\ (7,\pm2),\ (27/5,\pm54/25),
 \quad\text{or }p=p_*,
 \qquad p_*^3-17p_*^2+107p_*-243=0.
 \label{eq:critical-strip-alternatives}
\end{equation}
For $p=27/5$, $X=Y=Z=9/5$.

Finally, for any fixed column $c$, orthogonality gives
\begin{equation}
 \sum_{C\ni c}p_C=54,\qquad
 \sum_{C\ni c}J_C=0,\qquad
 \sum_{C\ni c}X_C^2\ge54.
 \label{eq:critical-anchored-sums}
\end{equation}
The last sum is $54+|\sum_j u_j^2|^2$ for the overlap phases $u_j$ of two rows.
Also, $\sum_{C\ni c}|\det E_C|^2=108$ yields the middle equality.
The cubic for $p_*$ has positive derivative and no root modulo five, so its unique real root is irrational and the first equality excludes it.
Writing the middle sum as $2A+(54/25)B=0$ with $|A|,|B|\le10$ gives $A=B=0$.
If $n_5,n_7,n_e$ count the ten corners with $p=5,7,27/5$, then
$25n_5+35n_7+27n_e=270$, $n_7,n_e$ are even and their total is at most ten.
Only $(0,0,10)$ and $(8,2,0)$ remain.
The first gives $\sum X_C^2=162/5<54$.
In the second, the two $p=7$ corners $C_1,C_2$ have opposite $J$.
A column belonging to $C_1$ but not to $C_2$ lies in $C_1,C_2^c$, whose equal nonzero $J$ values contradict the same count at that column.
This exhausts the alternatives. Equation~\eqref{eq:fourier-family-max-R} proves the converse.
\end{proof}

\subsubsection{Proof of both theorems}
\label{sec:proof-theorems-1-2}
\begin{proof}[Proof of Theorems~\paperref{thm:reconstruction} and~\paperref{thm:connected}]
\phantomsection\label{sec:non-tao-space}
Exchanging both block rows and block columns replaces
$\left(\begin{smallmatrix}E&B\\C&D\end{smallmatrix}\right)$
by $\left(\begin{smallmatrix}D&C\\B&E\end{smallmatrix}\right)$.
The new corner is strict, since $R(D)=4(\Imag\Psi(C))^2=R(\theta)>0$, and the new $B$ has the opposite imaginary-part sign. Uniqueness therefore represents the same class with the $H_+$ choice after changing the seed. Thus both choices of conjugate roots reach the same classes as the seed varies; their completions at a fixed seed need not be equivalent.

\phantomsection\label{sec:tao-isolation}
\phantomsection\label{proof:non-tao-connectedness}
The continuous image $f_+(\Omega)$ is path connected, and Eq.~\eqref{eq:methods-fourier-labels} joins it to $[F_6]$. Tao's isolated class, distinct from $[F_6]$~\cite[Lemma~3.3 and Secs.~5.6.1, 5.6.4]{TadejZyczkowski2006}, is absent. Propositions~\ref{prop:critical} and~\ref{prop:nonmatched} and the block exchange include every other class. Thus $f_+(\Omega)\cup\{[F_6]\}=\mathcal X$, proving Theorems~\paperref{thm:reconstruction} and~\paperref{thm:connected}.
\end{proof}

\subsection{The phase-ball cover and the exclusion of quartets}
\label{sec:finite-cover}
This section supplies the local half of the proof. The seed balls and the ball at $F_6$ are closed regions of classes; we show that their union is open in the phase distance, so that by Theorem~\paperref{thm:connected} it is the whole non-Tao space, and that no point of it is a quartet edge. Section~\ref{sec:phase-distance-cover} states the result, Sec.~\ref{sec:phase-ball-proof} encloses the reconstructed matrices on a seed ball, Sec.~\ref{sec:fourier-chart-proof} gives the chart at $F_6$, Sec.~\ref{sec:vector-exclusion} the vector covers and pair tests that exclude quartets, and Sec.~\ref{sec:boundary-transfers} the boundary transfers and the closed subdivisions used later for triples.

\subsubsection{Phase distance and the seed cover}
\label{sec:phase-distance-cover}
\phantomsection
\label{sec:phase-distance}
For real phase matrices, with entrywise exponentials and phases in radians,
\begin{equation}
 d_{\rm ph}([H],[K])=
 \inf_{\substack{e^{{\rm i}Q_H}\sim H\\e^{{\rm i}Q_K}\sim K}}\|Q_H-Q_K\|_F.
 \label{eq:methods-phase-distance}
\end{equation}
Equivalence allows row and column rephasings and permutations. We use Euclidean vector and operator norms and list phase entries row by row. The projection onto the space $V$ of matrices with zero row and column sums, and the dephased-corner map, are
\begin{equation}
 h(X)=\left(X_{jk}-\tfrac16\sum_mX_{jm}-\tfrac16\sum_mX_{mk}
                  +\tfrac1{36}\sum_{m,n}X_{mn}\right)_{j,k=1}^6,
 \label{eq:phase-projection}
\end{equation}
\begin{equation}
 L(Q)=(Q_{ab}-Q_{a1}-Q_{1b}+Q_{11})_{a,b=2,3},\qquad Lh=L.
 \label{eq:methods-phase-gauge-projection}
\end{equation}
\phantomsection\label{sec:figure-coordinates}
\begin{proposition}[Exclusion on the seed cover; computer-assisted]\label{prop:finite}
Let $W$ be the union of the images of the stored closed seed balls and the closed parameter ball through $F_6$, including their conjugates, transposes and conjugate transposes. In the phase distance of Eq.~\eqref{eq:methods-phase-distance}, $W$ is open, closed, nonextendable, and contains $F_6$ in a patch interior.
\end{proposition}
\begin{proof}
Compactness makes $W$ closed. The checks below establish exclusion, openness of interior images and transfer of every boundary class into an interior image. Hence $W$ is open and connectedness makes it the non-Tao space. Conjugation preserves overlaps; changing to the $H/\sqrt6$ basis and conjugating gives $H^T$, so exclusion respects the stated symmetries.
\end{proof}

\subsubsection{Completion enclosures on a seed ball}
\phantomsection\label{sec:phase-ball-proof}\label{sec:contraction-test}
For a square system $F$ and fixed square $Y$, the uniform bounds~\cite{Krawczyk1969}
\begin{equation}
 \|YF(0)\|\le b,\qquad
 \sup_{\|z\|\le\rho}\|I-YDF(z)\|\le\kappa<1,\qquad b+\kappa\rho<\rho
 \label{eq:common-contraction}
\end{equation}
make $z\mapsto z-YF(z)$ a contraction into the ball interior. The derivative bound implies that $Y$ is invertible, so the fixed point is the unique zero. Uniform bounds give continuous parameter dependence, and joint analyticity gives analytic dependence. This criterion supplies both completion methods below.

The $15$ complex row overlaps give $30$ real equations; row norms are already six. Their derivative and Taylor bounds are
\begin{equation}
 (\Phi_P(\delta))_{ab}=\sum_ke^{{\rm i}(P_{ak}-P_{bk}+\delta_{ak}-\delta_{bk})},\qquad 1\le a<b\le6,
 \label{eq:methods-phase-overlaps}
\end{equation}
\begin{equation}
 (J_Pv)_{ab}={\rm i}\sum_ke^{{\rm i}(P_{ak}-P_{bk})}(v_{ak}-v_{bk}),
 \label{eq:methods-phase-jacobian}
\end{equation}
\begin{equation}
 \|D^2\Phi\|\le2\sqrt3,\quad \|D^3\Phi\|\le2\sqrt6,\quad
 \|\Phi_P(\delta)-\Phi_P(0)-J_P\delta\|\le\sqrt3\|\delta\|_F^2,
 \label{eq:completion-derivatives}
\end{equation}
where $\Phi(q)=\Phi_0(q)$; differentiation and Cauchy--Schwarz prove the bounds.
At an exact center $q_0\in V$, choose $LZ=I_4$ and a basis $X$ of $V\cap\ker L$, of dimension $21$. With $J=D\Phi(q_0)$, verify $(JX)^TJX$ invertible and set
\begin{equation}
 \begin{aligned}
 W_1&=((JX)^TJX)^{-1}(JX)^T,& U_*&=(Z-XW_1JZ)M_{\theta_0},\\
 H_2&=-XW_1D^2\Phi(q_0)[U_*,U_*],&q_p(w)&=q_0+U_*w+\tfrac12H_2[w,w].
 \end{aligned}
 \label{eq:completion-predictor}
\end{equation}
For $G(w,z)=W_1\Phi(q_p(w)+Xz)$, $W_1JX=I$, $G(w,0)$ vanishes through second order, and $D_zG(0,0)=I$. Contraction on the whole parameter ball follows if
\begin{equation}
 \sup_{\|w\|\le r_H}\|G(w,0)\|
 +\tau\sup_{\substack{\|w\|\le r_H\\\|z\|\le\tau}}\|D_zG(w,z)-I\|<\tau.
 \label{eq:completion-contraction}
\end{equation}
The phases $q_p(w)+Xz(w)$ have seed $\theta_0+M_{\theta_0} w$. Near zero, uniqueness identifies them with the analytic Hadamard reconstruction, so all $30$ equations vanish although only $21$ combinations were imposed. The real analytic identity theorem extends this equality to the connected interior, and continuity includes the boundary. Rounding to stored $P,U$ gives $Q=P+Uw+\nu$, $\|\nu\|_F\le n$, provided
\begin{equation}
 n\ge\tfrac12\|H_2\|r_H^2+\|X\|\tau+\|q_0-P\|_F+r_H\|U_*-U\|.
 \label{eq:completion-remainder}
\end{equation}

\paragraph*{Enclosure on boxes.}
Alternatively cover the closed parameter ball by closed boxes $\mathcal B$. For the six normalized phases of $B$, let $F(\phi,w)$ be the real and imaginary off-diagonal entries of $S(\phi)S(\phi)^\dagger-G_h(\theta(w))$, ordered as in $\gamma$. Around a polynomial predictor $\widetilde\phi(w)$ verify, entrywise,
\begin{equation}
 |YF(\widetilde\phi,w)|\le b,\quad
 |I-YD_\phi F(\widetilde\phi+z,w)|\le A\quad(w\in\mathcal B,\ |z|\le\rho),
 \qquad b+A\rho<\rho,
 \label{eq:box-component-contraction}
\end{equation}
with $A,b\ge0$, $\rho>0$. This is Eq.~\eqref{eq:common-contraction} in the norm $\max_j|z_j|/\rho_j$. Apply it also to $C^T,G_v$. For a polynomial $\widetilde D$ and approximate inverse $A_0$,
\begin{equation}
 \begin{aligned}
 D-\widetilde D&=-(CE^\dagger+\widetilde D B^\dagger)(B^\dagger)^{-1},\\
 (B^\dagger)^{-1}&=\sum_{\ell\ge0}(I-A_0B^\dagger)^\ell A_0,
 \qquad\sup_{\mathcal B}\|I-A_0B^\dagger\|_\infty<1.
 \end{aligned}
 \label{eq:box-forced-block-error}
\end{equation}
The maximum absolute row-sum norm bounds the geometric tail. For $|z|<1/2$,
$|\arg(1+z)-\Imag(z-z^2/2)|\le |z|^3/[3(1-|z|)]$.
Integer phase shifts, projection by $h$, and interval polynomial bounds including rounding then verify
\begin{equation}
 \sup_{w\in\mathcal B,\,\|w\|\le r_H}\|Q_{\mathcal B}(w)-P-Uw\|_F\le n.
 \label{eq:box-completion-enclosure}
\end{equation}
Both constructions verify the fixed opposite nonzero signs of $\Imag\Psi(B),\Imag\Psi(C)$. Propositions~\ref{prop:sides} and~\ref{prop:matched} identify the completion; nearby Hadamards retain an interior seed and these signs, proving openness of interior images. The column and full phase bounds are
\begin{equation}
 r=\max_c\|U_c\|r_H+n,\qquad r_F=\|U\|r_H+n,
 \label{eq:local-normal}
\end{equation}
where $U_cw$ is column $c$ of $Uw$.

\subsubsection{Fourier neighborhood}
\label{sec:fourier-chart-proof}
At $F_6$ the side Gram Jacobian is singular, with rank five. Replace the fourth corner phase by $y=-2\Imag\Psi(B)$, whose sign selects the matched choice, and use
\begin{equation}
 \theta^0=(1,2,2,4)\pi/3,\qquad
 \xi=(\theta_1-\theta_1^0,\theta_2-\theta_2^0,\theta_3-\theta_3^0,y),\qquad R(E(\theta))=y^2.
 \label{eq:fourier-chart-coordinates}
\end{equation}
Since $\partial_4R(\theta^0)=1152\sqrt3\ne0$, the last equation locally recovers $\theta_4$. With $\xi=(u_1,u_2,u_3,y)$, $\phi^0=(3,4,5,0,2,4)\pi/3$ and $B=S(\phi^0+\varphi_h)$, solve the seven real equations
\begin{equation}
 BB^\dagger=G_h(\theta),\quad \Imag\Psi(B)=-y/2,\quad
 \theta=\theta^0+(u_1,u_2,u_3,t)
 \label{eq:fourier-augmented-system}
\end{equation}
for $(t,\varphi_h)$, and the corresponding system with $(u_1,u_3,u_2,-y)$ for $(t',\varphi_v)$ of $C^T$. Contraction on the stored ellipsoid $D_F=\{M_Fw:\|w\|\le r_H\}$, with $r_H\simeq0.13$, verifies both solutions, $g>0$ and $\partial_4R>0$ between the resulting phases. Thus $R=y^2$ implies $t=t'$. Enclosures on overlapping boxes lie in each other's uniqueness regions, so they agree and contain the exact Fourier solution at zero. The forced-block formula gives an analytic $H_F(\xi)$ in the interior, continuous on the closure. Conjugate roots prove the Hadamard equations for $y\ne0$; continuity proves them at $y=0$.

Choose continuous horizontal phases $Q(\xi)$ on this convex domain. For the orthogonality derivative $J_F$ at $F_6$, $T_F=\ker J_F\cap V$ has dimension four~\cite[Theorem~5.2]{TadejZyczkowski2008Defect} and the least singular value on $V\cap T_F^\perp$ is $\sqrt3$. The checks give $\|Q(\xi)-Q(0)\|_F\le0.1636<R_G=0.45$, a singular-value bound $1.7>2\sqrt3R_G$, and invertibility of $D_\xi\operatorname{proj}_{T_F}(Q(\xi)-Q(0))$. Two Hadamard displacements of norm at most $R_G$ with the same projection satisfy, by Eq.~\eqref{eq:completion-derivatives},
\begin{equation}
 0\ge(1.7-2\sqrt3R_G)\|\delta_1-\delta_2\|_F.
 \label{eq:fourier-projection-uniqueness}
\end{equation}
They coincide; the inverse function theorem therefore puts every nearby Hadamard in the chart, proving openness also at $F_6$. Overlap phase lifts agree because their dephased differences are integer multiples of $2\pi$ bounded by $4(0.1636)<2\pi$.
For direct quartet exclusion, the sharper residual bound $\sqrt2\|\delta\|_F^2$ gives the normal component $\|\nu\|_F\le\sqrt{2/3}\rho^2$. At $\rho=7/40$, the vector and graph tests below pass with $n=2501/100000$, $r=12606/100000$, including rounding. Since $0.1636<7/40$, they exclude quartets throughout the chart.

\subsubsection{Complete vector covers and pair tests}
\label{sec:vector-exclusion}
For the unbiasedness equations $f_k(\phi,Q)$ of Methods Eq.~\papereqref{eq:methods-vector-equations}, fix $\phi_1=0$ and cover the other phases by closed boxes. At a box center $u$ and matrix phases $P$, put $A_\lambda=\nabla_{\phi_2,\ldots,\phi_6}(\lambda\cdot f)$ and $B_\lambda=\nabla_Q(\lambda\cdot f)$. For half-width $a$ radians, Taylor expansion with $Q-P=Uw+\nu$ gives the necessary inequality
\begin{equation}
 |\lambda\cdot f(u,P)|\le a\|A_\lambda\|_1+r_H\|U^\dagger B_\lambda\|_2+n\|B_\lambda\|_F+E_\lambda.
 \label{eq:methods-vector-cut}
\end{equation}
Here $E_\lambda$ bounds the remainder uniformly in both enclosures, and $U^\dagger$ is the adjoint. A strict violation for any real $\lambda$ excludes the entire box.

For the explicit bounds, use cycles: $e(t)=e^{2\pi {\rm i}t}$, $(\widehat w,\widehat\nu,\widehat r_H,\widehat n,\widehat r,\widehat r_F)=(w,\nu,r_H,n,r,r_F)/(2\pi)$ and $\widehat\delta=(Q-P)/(2\pi)=U\widehat w+\widehat\nu$; $P$ remains in radians. Cover the five phases~\cite{Grassl2004,BrierleyWeigert2009,FuhrRzeszotnik2015,AnderssonBengtsson2017} by boxes of side $1/M$, $M=12,24,48,96$, retaining all $32$ children of each split unless excluded. Include the first phase's zero displacement and subtract the mean. Checking the $32$ vertices gives, for $a=1/(2M)$,
\begin{equation}
 \|x\|\le X=\sqrt{29/6}\,a,\qquad \|x\|_1\le5a,\qquad\|x\|_\infty\le3a/2.
\end{equation}
Set $q_{ji}=e(u_j)e^{-{\rm i}P_{ji}}/\sqrt6$, $S_i=\sum_jq_{ji}$, $S'_i=\sum_jq_{ji}e(x_j-\widehat\delta_{ji})$ and $K\ge\|e^{{\rm i}P}/\sqrt6\|_{\rm op}$. Since unbiasedness requires $|S_i'|=1$, the bounds
\begin{equation}
 |S'_i-S_i|\le2\pi(5a/\sqrt6+\widehat r),\qquad \|S'-S\|_2\le2\pi(KX+\widehat r_F)
\end{equation}
exclude a box when $||S_i|-1|$ or $\|(|S_i|-1)_i\|_2$ exceeds the corresponding right side. These follow by applying $|e(t)-1|\le2\pi|t|$ first to $x$, then to $\widehat\delta$.
For the sharper weighted test set
$c_i=(|S_i|^2-1)/(4\pi)$, $J_{ij}=-\Imag(\bar S_iq_{ji})$, $B_{ik}=\sum_jJ_{ij}U_{(j,i),k}$, $v_i=S_iy_i$ and $G_y=\max_j|\sum_i y_i\bar S_iq_{ji}|$. Every solution satisfies
\begin{equation}
 |y\cdot c|\le a\sum_{j>1}|(J^Ty)_j|+\widehat r_H\|B^Ty\|+\widehat n\|\diag(y)J\|_F+\mathcal E_0(y),
 \label{eq:local-linear}
\end{equation}
\begin{equation}
\begin{aligned}
 \mathcal E_0(y)&=\pi\Bigl[\min\{\|y\|_\infty(KX+\widehat r_F)^2,\|y\|_1(KX+\widehat r)^2\}
 +\min\{G_yX^2,\tfrac32aK\|v\|_2X\}\\
 &\hspace{15mm}+\tfrac{2X}{\sqrt6}\min\{\|v\|_2\widehat r_F,\|v\|_1\widehat r\}
 +\tfrac1{\sqrt6}\min\{\|v\|_\infty\widehat r_F^2,\|v\|_1\widehat r^2\}\Bigr].
\end{aligned}
 \label{eq:local-remainder}
\end{equation}
Expanding $|S_i'|^2-1$ with $|e(t)-1-2\pi {\rm i}t|\le2\pi^2t^2$ gives the squared increment and the $x$-only, mixed and $\widehat\delta$-only remainders. Cauchy--Schwarz with $\widehat r_F$ jointly or $\widehat r$ columnwise gives the alternatives. Substitute $U\widehat w+\widehat\nu$ for the matrix-error terms; for $y=e_i$, replace these by $\widehat r(\sum_jJ_{ij}^2)^{1/2}$. Row sums of $J$ vanish, justifying mean subtraction. Finally $f=4\pi c$ gives Eq.~\eqref{eq:methods-vector-cut} in radians with $E_\lambda=4\pi\mathcal E_0(\lambda)$.

The pair overlap in radians is
\begin{equation}
 \langle v(\phi),v(\psi)\rangle=\tfrac16\sum_je^{{\rm i}(\psi_j-\phi_j)}.
 \label{eq:methods-pair-overlap}
\end{equation}
Group retained boxes into labels, each a mean-zero phase ball of radius $s<0.1$ cycles. A box belongs to a label only if its center distance, minimized over integer windings and mean subtraction, plus $X$ is at most $s$. Within a label the unscaled overlap is at least $6-4\pi\sqrt6s>\sqrt6$, excluding both orthogonality and unbiasedness.
For labels $A,B$, put $Q_j=e(u_j-v_j)$, $S=\sum_jQ_j$, $r_e=s_A+s_B$. Orthogonality requires $|S|\le2\pi\sqrt6r_e$; unbiasedness requires $\bigl||S|-\sqrt6\bigr|\le2\pi\sqrt6r_e$. The same Taylor inequalities give the following residual, derivative and remainder triples:
\begin{equation}
 q=\tfrac1{2\pi}\binom{\Real S}{\Imag S},\quad
 D=\begin{pmatrix}-\Imag Q_1&\cdots&-\Imag Q_6\\\Real Q_1&\cdots&\Real Q_6\end{pmatrix},\quad
 \epsilon=\pi r_e^2 \quad\text{(orthogonality)},
 \label{eq:local-edges}
\end{equation}
\begin{equation}
 q=\tfrac{|S|^2/6-1}{4\pi},\quad D=-\tfrac16(\Imag(\bar SQ_j))_{j=1}^6,\quad
 \epsilon=\pi r_e^2(1+|S|/6)\quad\text{(unbiasedness)}.
 \label{eq:unbiased-pair-test}
\end{equation}
Combining these equations with each vector's unbiasedness equations gives, for real weights,
\begin{equation}
 \begin{aligned}
 |y_A\cdot c^A+y_B\cdot c^B+y_e\cdot q|
 \le{}&s_A\|J_A^Ty_A+D^Ty_e\|+s_B\|J_B^Ty_B-D^Ty_e\|\\
 &+\widehat r_H\|B_A^Ty_A+B_B^Ty_B\|
 +\mathcal E_A^{\rm ball}(y_A)+\mathcal E_B^{\rm ball}(y_B)+\epsilon\|y_e\|.
 \end{aligned}
 \label{eq:local-joint}
\end{equation}
Here $\mathcal E^{\rm ball}=\widehat n\|\diag(y)J\|_F+\mathcal E_0(y)$ with $X=s$ and the $x$-only remainder replaced by $G_ys^2$. The common Hadamard motion is combined before taking norms. Setting $y_A=y_B=0$ gives the separate pair tests; $y_e=(\Real S,\Imag S)$ gives the orthogonality test. Only impossible edges are removed, and a quartet uses at most one vector per label. Absence of the main-text graph pattern therefore excludes quartets on the whole closed ball.

\subsubsection{Boundary transfers and closed subdivisions}
\label{sec:boundary-transfers}
Each parameter sphere $\|w\|_2=r_H$ is covered by spherical caps; an exact polar-facet certificate, checked in integers, shows that the caps cover the sphere. For a source cap and strict target $t$, verify agreement of both imaginary-part signs and, for some $N\in\mathbb Z^4$,
\begin{equation}
 \sup_{w\in\mathrm{cap}}\|M_t^{-1}(\theta_s(w)-\theta_t^0+2\pi N)\|_2<r_{H,t}.
 \label{eq:local-transfer}
\end{equation}
The bound holds on the whole cap by a $5\times5$ positive-definite S-lemma witness for the affine predictor of the source ball's enclosure, which accounts for its remainder; after a corner change use $L$ to extract the seed. Reconstruction uniqueness identifies the classes. For a Fourier target, the stored frame satisfies $D_F\subset E_{1/5}$, where
\begin{equation}
 E_\rho=\{(u_1,u_2,u_3,y):u_1^2-u_1(u_2+u_3)+\tfrac74(u_2^2+u_3^2)+\tfrac12u_2u_3+(y/24)^2\le\rho^2\},\qquad 0\le\rho\le1/5.
\end{equation}
Throughout the cap we verify that the signed parameter $\xi'$ of Eq.~\eqref{eq:fourier-chart-coordinates} for the transformed source $K$ lies in the interior of $D_F$ and that $d_{\rm ph}([K],[F_6])<3/20$. The Fourier transfer theorem then gives $K\sim H_F(\xi')$ without a separate side or orientation test.

\phantomsection\label{sec:tc-subdivision}
For local triple certification on a seed ball, split the enclosing cube $[-r_H,r_H]^4$ into all sixteen closed children. A cube $\prod_i[a_i,b_i]$ is outside the original ball only when $\sum_i\operatorname{dist}(0,[a_i,b_i])^2>r_H^2$; tangencies are retained. Every other terminal cube must lie in a certified local parameter ball, checked at all sixteen vertices of the affine coordinate change or by a bound for their maximum. Each local normal form is
\begin{equation}
 Q_i(w_i)=p_i(w_i)+X_i z_i(w_i),\quad
 p_i(w_i)=Q_{0i}+U_iw_i+\tfrac12H_{2i}[w_i,w_i],\quad \|z_i\|\le\tau_i.
 \label{eq:tc-normal-form}
\end{equation}
Containment must also identify the original analytic branch. At a seed interior to parent $p$ and child $c$, align the lifted seeds using a peripheral row/column permutation $\Pi$ preserving the corner and an integer phase matrix $N$. Verify
\begin{equation}
 \|X_p^+(\Pi p_c(w_c)+2\pi h(N)-p_p(w_p))\|
 +\|X_p^+\Pi X_c\|\tau_c<\tau_p,\qquad X_p^+=(X_p^TX_p)^{-1}X_p^T.
 \label{eq:tc-branch-overlap}
\end{equation}
Equal lifted seeds put the difference in $\operatorname{ran}X_p$; the inequality puts the child's actual Hadamard in the parent's uniqueness tube. The reconstructions agree near this seed, hence throughout their connected common interior by the analytic identity theorem, and on its boundary by continuity. At a record's own center its correction is zero; parent and child may be interchanged. For boundary-only contact, repeat the parent's contraction on a slightly larger auxiliary ball and identify its branch by the same argument. This changes no original domain. These rules cover every original boundary point and apply also to leaves using the original local classifier.

\subsection{Proof of the classification of all MUB sets}
\label{sec:tc-classification}
On $2{,}771$ seed balls and the Fourier chart every triple is classified; on the other four, the triples whose first transition minimizes the score $\Gamma_F$ of Eq.~\eqref{eq:gamma-score} among their three transitions, ties included, are classified. Lemma~\ref{lem:minimum-transition} shows that this suffices to classify every triple. On each piece of the cover, a pattern is excluded, classified by the order-three symmetry, or discharged by a strict comparison that rules out a minimizing first transition. The symmetric triples are exactly the two known types. Section~\ref{sec:tc-patterns} sets up the patterns and the triple equations. Sections~\ref{sec:tc-taylor}--\ref{sec:tc-weak-normal-form} give the tests that exclude a pattern or force its symmetry: column tubes, basis exclusions, the symmetry criterion, and the normal-form localisation near the family. Section~\ref{sec:tc-recognition} solves the symmetric triples exactly, Sec.~\ref{sec:tc-crossings} treats the two crossings in the Fourier chart, and Sec.~\ref{sec:tc-verification} states the certificate and proves the theorem. Section~\ref{app:triple-seeds} gives the seeds of every triple.

\restated{Theorem~\paperref{thm:all-mubs} (Classification of all MUB sets; restated; computer-assisted)}{Up to equivalence, every nonempty set of MUBs in $\mathbb C^6$ is a single basis $\{I_6\}$; a pair $\{I_6,H/\sqrt6\}$, where $H$ is Tao's matrix, $F_6$, or $H_+(\theta)$ for a strict seed $\theta$; or a triple in the continuous two-parameter family, or the isolated Fourier triple. No set contains four or more bases.}

\subsubsection{Patterns and the triple equations}
\label{sec:tc-patterns}
We use the same $2{,}775$ seed balls and Fourier chart, with their original centers, frames and radii. Each ball's radius $r_H$ bounds its framed four seed coordinates, not all entry phases. Triple equivalence allows a common ambient unitary, independent column phases and permutations, and permutation of the three bases.
The complete vector cover retains every column of every completing $K$. Labels contain no orthogonal pair, so enumeration of every six-clique retains every basis. If labels are grouped into colors, each color must be an independent set of this necessary orthogonality graph. A six-column choice is a \emph{pattern}; it may be phantom because pairwise compatibility need not be realizable at one common $H$. Deleting labels lacking a compatible label in another required color preserves actual bases by induction; the remaining enumeration must be complete.

Write $H_{rc}=e^{{\rm i}Q_{rc}}$, $K_{rc}=e^{{\rm i}V_{rc}}$, $x=(Q,V)\in\mathbb R^{72}$, and define
\begin{equation}
 \begin{aligned}
 O^H_{ab}&=\sum_re^{{\rm i}(Q_{rb}-Q_{ra})},&O^K_{ab}&=\sum_re^{{\rm i}(V_{rb}-V_{ra})},\quad a<b,\\
 F^{\rm mix}_{ab}&=\frac{|\sum_re^{{\rm i}(V_{rb}-Q_{ra})}|^2-6}{6},&
 F&=(\Real O^H,\Imag O^H,\Real O^K,\Imag O^K,F^{\rm mix}).
 \end{aligned}
 \label{eq:tc-equations}
\end{equation}
Thus $F$ has $96$ real components, and $F=0$ exactly characterizes the triple $(I_6,H/\sqrt6,K/\sqrt6)$. All phases below are in radians. Each pattern is excluded, classified by symmetry, or discharged by the strict ordering test below.

\paragraph*{Ordering the three transitions.}
For a real phase array $Q$, define the $225$ plaquette differences
\begin{equation}
 (\mathcal RQ)_{ik;jl}=Q_{ij}+Q_{kl}-Q_{il}-Q_{kj},\qquad i<k,\ j<l.
\end{equation}
Let $a,b$ range over permutations of the labels $0,\ldots,5$, assigned to the six rows and columns, with $a_1=b_1=0$. Write $v(a,b)_{ik;jl}=(a_i-a_k)(b_j-b_l)$ and let $\operatorname{dist}_{2\pi}$ take the distance of each entry to $2\pi\mathbb Z$. The score used throughout the certificates is
\begin{equation}
 \Gamma_F(e^{{\rm i}Q})=\frac1{15}\min_{a,b}
 \left\|\operatorname{dist}_{2\pi}\!\left(\mathcal RQ-\frac\pi3v(a,b)\right)\right\|_2.
 \label{eq:gamma-score}
\end{equation}
There are $120^2=14{,}400$ label pairs. Plaquettes remove row and column phases. Permutations relabel the minimum; translating all row or column labels modulo six restores $a_1=b_1=0$ without changing a circular distance. Negating one set of labels and interchanging the two sets proves invariance under conjugation, transpose and adjoint. Thus $\Gamma_F$ is independent of the representative and of the direction of a transition. Since $\|\mathcal RX\|_2=6\|h(X)\|_F$ and circular distance is $1$-Lipschitz,
\begin{equation}
 |\Gamma_F(e^{{\rm i}Q})-\Gamma_F(e^{{\rm i}P})|\le\frac25\|h(Q-P)\|_F.
 \label{eq:gamma-lipschitz}
\end{equation}

\begin{supplemma}[Minimum transition]\label{lem:minimum-transition}
Suppose the cover classifies every triple $(I_6,H/\sqrt6,K/\sqrt6)$ whose transition $H$ satisfies
\begin{equation}
 \Gamma_F(H)\le\Gamma_F(K),\qquad
 \Gamma_F(H)\le\Gamma_F(H^\dagger K/\sqrt6).
 \label{eq:minimum-transition}
\end{equation}
Then it classifies every triple. Ties are included.
\end{supplemma}
\begin{proof}
Choose a pair of bases whose transition minimizes the same score among the three pairs. Order this pair first and apply the adjoint of its first basis to all three bases. The resulting $H$ satisfies Eq.~\eqref{eq:minimum-transition}, since reversing a transition takes its adjoint and preserves the score. Tao's class admits no third basis, so $H$ lies in the non-Tao cover. The assumed classification applies. The conjugation and transpose transfers of the cover preserve both recognized types, as shown in Sec.~\ref{sec:tc-recognition}. They also preserve Eq.~\eqref{eq:minimum-transition}: each permutes the three scores while fixing the first, since conjugation maps $(H,K,L)$ to $(\bar H,\bar K,\bar L)$ and exchanging the first two bases maps it to $(H^\dagger,L,K)$.
\end{proof}

For a given pattern, choose $M\in\{K,H^\dagger K/\sqrt6\}$. Certify a lower bound $g_H$ for $\Gamma_F(H)$ and an upper bound $g_M$ for $\Gamma_F(M)$ throughout that pattern. An ordering discharge requires
\begin{equation}
 g_M<g_H.
 \label{eq:ordering-discharge}
\end{equation}
Then no triple in the pattern can satisfy Eq.~\eqref{eq:minimum-transition}. If $H$ ties for the minimum, the strict inequality cannot hold. A center lower bound minimizes over all $14{,}400$ alignments; a center upper bound may use any one alignment. Equation~\eqref{eq:gamma-lipschitz} extends them to the whole phase enclosure. For $L=H^\dagger K/\sqrt6$, first enclose this product using the same $H$ and column tubes, bound its arguments on disks avoiding zero, and then apply the score bound. All comparisons and enclosure errors must be certified. No Fourier-neighborhood classification threshold is used in this reduction.

\subsubsection{Uniform phase bounds and column tubes}
\label{sec:tc-taylor}
\begin{supplemma}[Fourier bounds]\label{lem:tc-fourier}
For a finite Fourier sum, derivatives are obtained termwise, and for real $t$
\begin{equation}
 \left|e^{{\rm i}t}-\sum_{j=0}^m\frac{({\rm i}t)^j}{j!}\right|\le\frac{|t|^{m+1}}{(m+1)!}.
 \label{eq:tc-fourier-tail}
\end{equation}
If a phase increment along a segment is bounded by $(1-s)\beta_0+s\beta_1$, its integrated quadratic remainder is at most $(\beta_0^2+\beta_0\beta_1+\beta_1^2)/6$.
\end{supplemma}
Both statements follow by integrating the Taylor remainder and using $|e^{{\rm i}t}|=1$. Overlaps in Eq.~\eqref{eq:tc-equations} have two-entry frequencies; mixed components are $\frac13\sum_{r<s}\cos(V_{rb}-V_{sb}-Q_{ra}+Q_{sa})$. For mean-zero six-vectors $d$, the cubic remainder bounds are $\|d\|^3/6$ and $\sqrt2\|d\|^3/3$, respectively, since $\sum_{r<s}(d_r-d_s)^2=6\|d\|^2$ and $|d_r-d_s|\le\sqrt2\|d\|$. Signed terms sharing variables are combined before norms are taken.

Column enclosures share the same seed parameter $w$:
\begin{equation}
 V_c=v_c+T_cw+e_c,\qquad \mathbf1^Te_c=0,\qquad\|e_c\|\le\epsilon_c.
 \label{eq:tc-column-tube}
\end{equation}
Initially take the full union of retained bins; width $2\pi/M$ gives mean-zero radius $(\pi/M)\sqrt{29/6}$. For column unbiasedness, optionally augmented by orthogonality to enclosed columns, expand in orthonormal mean-zero coordinates as $f=f_0+J_ee+J_ww+J_nn+\mathcal R$. Any zero satisfies, for any real $A$,
\begin{equation}
 e=-Af_0-AJ_ww-AJ_nn+(I-AJ_e)e-A\mathcal R.
 \label{eq:tc-conditional-contraction}
\end{equation}
Enclosing this identity refines the tube; using all six errors refines the full basis. Iteration uses established bounds only, retaining residuals, inverse defects, center/carrier changes and reference-column errors. The common quadratic terms, including $J_HH_2[w,w]/2$, are combined; Lemma~\ref{lem:tc-fourier} bounds the tails.

If the tubes leave a pattern unresolved, each colour is replaced by its retained closed fine bins at $M=96$, of mean-zero radius $(\pi/96)\sqrt{29/6}$, and the pair tests of Sec.~\ref{sec:vector-exclusion} are repeated on pairs of bins with the same whole-ball enclosure. Arc consistency then removes a bin only if some other column has no compatible bin, so every bin of an actual basis survives, and an exhaustive six-clique search that finds none excludes the pattern.

\subsubsection{Basis exclusions and symmetry}
\label{sec:tc-joint}
For a full-basis test of $y^TF$, put $q=(Q_0,v_1,\ldots,v_6)$, $J=DF(q)$, $B=D^2(y^TF)(q)$, $\mathcal L=(U;T_1;\ldots;T_6)$ and $B_2=y^TJ_HH_2$, reshaped into a $4\times4$ matrix. Let $I_H,I_K$ insert the horizontal $H$ error and the six mean-zero column errors, and set $N_H=\|H_2\|r_H^2/2+\|X\|\tau$, $N_K=(\sum_c\epsilon_c^2)^{1/2}$. Expansion gives
\begin{equation}
 \begin{aligned}
 |y^TF(q)|\le{}&r_H\|y^TJ\mathcal L\|+\tau\|y^TJ_HX\|+\sum_c\epsilon_c\|y^TJ_{K_c}\|
 +\tfrac12r_H^2\|\mathcal L^TB\mathcal L+B_2\|\\
 &+r_H\sum_{a\in\{H,K\}}N_a\|\mathcal L^TBI_a\|
 +\tfrac12\sum_{a,b\in\{H,K\}}N_aN_b\|I_a^TBI_b\|+\mathcal R_3.
 \end{aligned}
 \label{eq:tc-joint-support}
\end{equation}
For each complex overlap, multiply $d^3/6$ by the norm of its two real weights; for each mixed equation multiply $\sqrt2d^3/3$ by its absolute weight, and sum for $\mathcal R_3$. Bound $d$ from the relevant column differences of the normal form and tubes, combining $U_b-U_a$, $T_b-T_a$ or $T_b-U_a$ before norms. The quadratic $J_HH_2$ term is likewise combined before bounding. A strict failure excludes the pattern. Replacing $y^T$ by each row of a left inverse proposal $A$ for the six columns gives the same bound for every component of the six column errors; their norms refine all six tubes together, as in Eq.~\eqref{eq:tc-conditional-contraction}.

\phantomsection\label{sec:tc-symmetry-proof}
Remove common row phases and independent column phases of $H,K$ with the rank-$55$ orthogonal projector
\begin{equation}
 W_{(b,r,c),(d,s,j)}=(\delta_{rs}-1/6)(\delta_{bd}\delta_{cj}-1/12),\qquad b,d\in\{H,K\}.
 \label{eq:tc-joint-gauge}
\end{equation}
Joint phase norms refer to this slice, with explicit integer $2\pi$ windings. A proposed physical symmetry fixes the rays of an anchor basis and permutes each other basis in two three-cycles, with eigenvalues $1,\omega,\omega^2$ each twice. Let $R,S,T$ be the row, $H$-column and $K$-column permutations; $P$ acts on phases by these permutations. For an identity anchor, $R=I$, $S,T$ have two three-cycles, and
$b^H_{rc}=-2\pi\ell_rp_S(c)/3$, $b^K_{rc}=-2\pi\ell_rp_T(c)/3$, with two row labels of each $0,1,2$ and $p$ advancing around each cycle. For an $H$ anchor, $S=I$, $R,T$ have two three-cycles and $b^H_{rc}=2\pi p_R(r)\ell_c/3$, $b^K=0$; exchange $H,K$ for a $K$ anchor.
The affine map $x\mapsto b+P(x-b)$ exponentiates to common row and independent column operations. It induces $F(b+P(x-b))=YF(x)$: on overlaps, permute pairs, conjugate reversed pairs and apply the cubic-root column phase; on mixed equations, permute indices. Thus $P^3=Y^3=I$, $Y^TY=I$, $PW=WP$, checked over $\mathbb Q(\sqrt3)$.

Put $G=(I+P+P^2)/3$, $E=(I-G)W$, $Q_Y=I-(I+Y+Y^2)/3$ and $A=EA_0Q_Y$ for any real inverse proposal $A_0$. For a covered zero, $\bar x=b+G(x-b)$ and $e=x-\bar x$ satisfy $Ee=e$ and $AF(\bar x)=0$ by covariance, although $\bar x$ need not solve $F=0$. Hence
\begin{equation}
 0=A\int_0^1DF(\bar x+se)e\,ds,\qquad
 \left\|E-A\int_0^1DF(\bar x+se)E\,ds\right\|\le q<1
 \quad\Longrightarrow\quad e=0.
 \label{eq:tc-symmetry}
\end{equation}
This is $\|e\|\le q\|e\|$, proving exact physical symmetry. To bound $q$, write $x=b+W(q_0-b)+W(\mathcal Lw+n)$, $\|n_H\|\le N_H$, $\|n_{K_c}\|\le\epsilon_c$. Set $a=b+GW(q_0-b)$, $m=EW(q_0-b)$, $C=(I+G)W/2$ and $\mathcal B(v)=AD^2F(a)[v,E\,\cdot]$. The mean segment displacement is $m/2+C(\mathcal Lw+n)$, giving
\begin{equation}
 D_0=\|\mathcal B(m/2)\|+r_H\|[\mathcal B(C\mathcal L_1)\ \cdots\ \mathcal B(C\mathcal L_4)]\|.
 \label{eq:tc-symmetry-linear}
\end{equation}
If $h,h_H,h_K$ bound the flattened signed map $n\mapsto\mathcal B(Cn)$ on the whole error space and its two blocks, take $D_1=\min\{h\sqrt{N_H^2+N_K^2},h_HN_H+h_KN_K\}$. In $F(x)=\Real\sum_\nu a_\nu e^{{\rm i}\ell_\nu\cdot x}$, bound endpoint phase increments from $a$ by $\beta_{0\nu},\beta_{1\nu}$. Apply $GW$ to the averaged endpoint's carrier and errors, and $W$ with constant displacement $m$ to the solution endpoint. The row support on the errors is
\begin{equation}
 \tfrac12r_H^2\|\ell_HH_2\|+\tau\|\ell_HX\|
 +\sum_c\|\ell_{K_c}(I-\mathbf1\mathbf1^T/6)\|\epsilon_c.
 \label{eq:tc-frequency-support}
\end{equation}
Lemma~\ref{lem:tc-fourier} bounds the integrated Jacobian remainder entrywise by
\begin{equation}
 M_{ij}=\sum_\nu|(Aa_\nu)_i|\,|(\ell_\nu E)_j|
 (\beta_{0\nu}^2+\beta_{0\nu}\beta_{1\nu}+\beta_{1\nu}^2)/6,
 \label{eq:tc-symmetry-tail}
\end{equation}
\begin{equation}
 q=\|E-ADF(a)E\|+D_0+D_1+\|M\|<1.
 \label{eq:tc-symmetry-q}
\end{equation}
Thus the test includes the inverse defect and every phase error on every averaging segment, without assuming $a$ or $\bar x$ is a solution.

\subsubsection{Normal-form localisation near the family}
\label{sec:tc-weak-normal-form}
Close to the continuous family the triple equations are nearly degenerate, and the tests above need very small cells. The following test localizes every solution in coordinates adapted to the degeneracy. Fix a cell and a pattern with its retained fine bins. Let $F$ now denote the $66$ real orthogonality and mixed unbiasedness equations as a function of the phases $q$ of $H$ and the column-centred phases $y$ of $K$, and write
\begin{equation}
 y=y^*+ET\zeta,\qquad G(\zeta)=BF(q^*,y^*+ET\zeta),\qquad \zeta=(\zeta_s,\zeta_w)\in\mathbb R^{28}\oplus\mathbb R^2,
 \label{eq:tc-weak-coordinates}
\end{equation}
Here $E\in\mathbb R^{36\times30}$ is the block Helmert isometry onto mean-zero column phases, $T\in\mathbb R^{30\times30}$ is invertible, and $B\in\mathbb R^{30\times66}$ is a left preconditioner; these local symbols are separate from the corner block and symmetry projectors used earlier. A Gauss--Newton search in $(u,y)$, with $q=Q_0+Uu$, proposes a reference $(q^*,y^*)$ near a triple of the family; the invertible frame $T$ and the left preconditioner $B$ are proposed from the singular value decomposition of $D_yF$ there, with the weak block $\zeta_w$ along its two smallest singular directions. No proposal is assumed to be exact.

For the $H$ displacement $u$ across the cell, $\|u\|\le r$, put $\zeta=z+\Lambda u$ with $\Lambda_w=0$ and define the cubic normal-form coordinate
\begin{equation}
 \begin{aligned}
 \tilde z_s&=z_s-q_2(z_w)-q_3(z_w),\qquad q_2(w)=-\tfrac12D^2_{ww}G_s(0)[w,w],\\
 q_3(w)&=-D^2_{sw}G_s(0)[q_2(w),w]-\tfrac16D^3_{www}G_s(0)[w,w,w].
 \end{aligned}
 \label{eq:tc-cubic-normal-form}
\end{equation}
The shear $(z_s,z_w)\mapsto(\tilde z_s,z_w)$ is invertible, so every zero is retained. The derivative at a proposed reference need not be the identity. Put $L=DG(0)-I$ and $N(\zeta)=G(\zeta)-G(0)-DG(0)\zeta$. At $u=0$, writing $w=z_w$ gives the exact strong equation
\begin{equation}
 \begin{aligned}
 0={}&G_s(0)+\tilde z_s
 +L_{ss}\bigl(\tilde z_s+q_2(w)+q_3(w)\bigr)+L_{sw}w\\
 &+N_s(\zeta)+q_2(w)+q_3(w).
 \end{aligned}
 \label{eq:tc-defect-decomposition}
\end{equation}
Only the pure weak terms of degrees two and three in $N_s+q_2+q_3$ cancel. The terms $L_{ss}q_2$, $L_{ss}q_3$, $L_{sw}w$, the strong linear defect and the corresponding weak-block defects remain. To define the slacks without omitting any terms, set
\begin{equation}
 \begin{aligned}
 \mathcal E(u,t,w)={}&BF\!\left(q(u),y^*+ET\bigl[(t+q_2(w)+q_3(w),w)+\Lambda u\bigr]\right)\\
 &-G(0)-(t,w).
 \end{aligned}
 \label{eq:tc-complete-remainder}
\end{equation}
For $a=s,w$, let $\sigma_a(s,w)$ be a non-decreasing certified upper bound for $\|\mathcal E_a(u,t,v)\|$ on the portion of the retained bin domain being tested, with $\|t\|\le s$, $\|v\|\le w$, $\|u\|\le r$, using the entire certified enclosure of $q(u)$, including its reconstruction error. Thus residuals, derivative defects, carrier errors and the motion of $H$ are all retained. Every zero with $\|\zeta\|\le R$ and $\|\zeta_s\|\le s_{\rm cap}$ satisfies
\begin{equation}
 \|\tilde z_s\|\le\|G_s(0)\|+\sigma_s(\|\tilde z_s\|,\|z_w\|),\qquad
 \|z_w\|\le\|G_w(0)\|+\sigma_w(\|\tilde z_s\|,\|z_w\|).
 \label{eq:tc-localisation}
\end{equation}
The bounds on Eq.~\eqref{eq:tc-complete-remainder} combine the defect terms above with the nonlinear terms and the Taylor tails of Lemma~\ref{lem:tc-fourier}. Homogeneous weak polynomials use the identity $\|(w_1^d,\ldots,\binom dj^{1/2}w_1^{d-j}w_2^j,\ldots,w_2^d)\|=\|w\|^d$. Derivative tensors use Gram flattenings across the strong and weak blocks. For the mixed second derivative, sixteen equally spaced weak directions modulo sign give an upper bound only after multiplication by $\sec(\pi/32)$, which covers the angles between them.

In particular, if $q(u)=q^*+Uu+\tfrac12H_2[u,u]+e_H(u)$, the pure quadratic carrier term before bounding is
\begin{equation}
 \begin{aligned}
 &\tfrac12BF_qH_2[u,u]+\tfrac12BF_{qq}[Uu,Uu]\\
 &\qquad+BF_{qy}[Uu,ET\Lambda u]+\tfrac12BF_{yy}[ET\Lambda u,ET\Lambda u],
 \end{aligned}
 \label{eq:tc-quadratic-carrier}
\end{equation}
with derivatives at $(q^*,y^*)$ and the $e_H$ contribution bounded separately. The mixed term has a positive sign. Signed coefficients are combined before norms are taken.

Let $W=\min(R,w_{\rm cap})$, where $w_{\rm cap}$ bounds the weak norm on the retained bins. From Eq.~\eqref{eq:tc-cubic-normal-form} and $\|q_j(v)\|\le\kappa_j\|v\|^j$ for $j=2,3$, the initial box for $(\|\tilde z_s\|,\|z_w\|)$ is
\begin{equation}
 [0,\min(R,s_{\rm cap})+\|\Lambda_s\|r+\kappa_2 W^2+\kappa_3 W^3]\times[0,W].
\end{equation}
For a box $[s_l,s_u]\times[w_l,w_u]$, use $\|\zeta_s\|\le\min(s_{\rm cap},\sqrt{R^2-w_l^2})$. Monotonicity bounds each slack on the retained domain in the box by its value at $(s_u,w_u)$. We therefore discard the box only if
\begin{equation}
 s_l>\|G_s(0)\|+\sigma_s(s_u,w_u)
 \quad\text{or}\quad
 w_l>\|G_w(0)\|+\sigma_w(s_u,w_u).
 \label{eq:tc-box-rejection}
\end{equation}
Otherwise it is retained or subdivided. Both comparisons are strict, so boundary points are retained.

The cap $s_{\rm cap}$ comes from the retained bins after their mean-zero phase displacements $y-y^*$ have been mapped to the $30$ coordinates $E^T(y-y^*)$. For any proposed weak point $w_0$, put $x_0=T(0,w_0)$ in those coordinates. Every point of the product of bins has $\|\zeta_s\|\le\|T^{-1}\|(\sum_cr_c^2)^{1/2}+\|T^{-1}x_0-(0,w_0)\|$, where $r_c$ is the largest distance of a transformed bin center from the corresponding five-coordinate block of $x_0$, plus its transformed bin radius. The last term is exactly zero for $T^{-1}$; retaining its outward-rounded enclosure is harmless. If some boxes survive, the retained bins are partitioned by their reach into a core and outer parts, recursively. Every six-clique lies in exactly one part, and arc consistency within a part preserves it. Each part is excluded by Eq.~\eqref{eq:tc-localisation}, emptied, or passed to the symmetry test with the per-column tubes
\begin{equation}
 \|e_c\|\le n_s\bigl(s_u+\kappa_2w_u^2+\kappa_3w_u^3\bigr)+n_ww_u
 \label{eq:tc-normal-form-tubes}
\end{equation}
about the moving reference $y^*+ET\Lambda u$, where $n_s,n_w$ bound the strong and weak column blocks of $T$ and $\kappa_2,\kappa_3$ bound $q_2,q_3$. There the symmetry test may also use the $u$-dependent preconditioner $(I-\sum_ku_kV_k)A$, with $V_kE=V_k$ chosen to cancel the part of $AM(u)E$ linear in $u$; with $\nu$ and $\mu$ the norms of $[V_1\ \cdots\ V_4]$ and $[V_jV_k]_{j,k}$, it gives $q=\mu r^2+(1+\nu r)q_0$, where $q_0$ bounds the other defects of Eq.~\eqref{eq:tc-symmetry-q}. A pattern passes this test only if every part is excluded or classified. All of its bounds are evaluated in outward-rounded interval arithmetic. A record using the cubic test must be replayed with Eq.~\eqref{eq:tc-defect-decomposition}, the positive mixed term in Eq.~\eqref{eq:tc-quadratic-carrier}, and the strict rule~\eqref{eq:tc-box-rejection}. The reported classification run used the cubic normal-form verifier on $151{,}156$ leaves across all $402$ refined seed-ball certificates.

\subsubsection{Algebraic recognition of the symmetric triples}
\label{sec:tc-recognition}
Move the anchor to the identity by a common unitary and order its eigenspaces as three coordinate pairs. With $t,s\in\mathbb Z_3$, $a,b\in\mathbb Z_2$, normalizing the column orbits gives
\begin{equation}
 H_{ta,sb}=\omega^{ts}(-1)^{ab}x_t^b,\qquad
 K_{ta,sb}=\omega^{ts}(-1)^{ab}y_t^b
 \begin{cases}u_t,&a=0,\\v_t,&a=1,\end{cases}
 \label{eq:tc-cyclic-coordinates}
\end{equation}
with twelve unit phases. Indeed Fourier-transforming orbit orthogonality in $s$ makes the generators orthogonal in each two-dimensional eigenspace; equal entry moduli give the signs $(-1)^{ab}$. Apply the same normalizing row phases to $H,K$. These formulas make both Hadamard; the remaining condition is flatness of $H^\dagger K/\sqrt6$.

The transition has four circulant $3\times3$ blocks. Divide by $\sqrt6$ and diagonalize their shifts to obtain unitary blocks
\begin{equation}
 C_t=\begin{pmatrix}\alpha_t&\beta_t\\z_t\bar\beta_t&-z_t\bar\alpha_t\end{pmatrix},\quad
 |z_t|=1,\qquad
 \alpha_t=\tfrac1{\sqrt6}\sum_sa_s\omega^{st},\quad
 \beta_t=\tfrac1{\sqrt6}\sum_sb_s\omega^{st},
 \label{eq:tc-fourier-blocks}
\end{equation}
where $a_s,b_s$ are the top circulants' unit entries. Put $\gamma=\sum_sa_{s+1}\bar a_s$, $q=|\gamma|^2$, $p=\Real(\gamma^3)$. Unitarity gives $\sum_sb_{s+1}\bar b_s=-\gamma$. The consecutive unit ratios have product one and polynomial $z^3-\gamma z^2+\bar\gamma z-1$; their squared Vandermonde is nonpositive real. Applying this to $a,b$ gives
\begin{equation}
 27-18q-q^2\pm8p\ge0,\quad q\le6\sqrt3-9<\tfrac32,\quad
 |\alpha_t|^2=\tfrac{3+2\Real(\gamma\omega^t)}6,\quad|\beta_t|^2=1-|\alpha_t|^2.
 \label{eq:tc-amplitude-bounds}
\end{equation}
Thus no $\alpha_t,\beta_t$ vanishes, including repeated-root boundaries. For $A_t=\bar\alpha_t\alpha_{t+1}$, $B_t=\bar\beta_t\beta_{t+1}$ and $w_t=z_t\bar z_{t+1}$, flatness gives
\begin{equation}
 \sum_tA_t=\sum_tB_t=0,\qquad
 \sum_tw_tA_t=\sum_tw_tB_t=0,\qquad\prod_tw_t=1.
 \label{eq:tc-recognition-equations}
\end{equation}
If $A,B$ are independent, their common kernel is $\operatorname{span}(1,1,1)$; hence $w_t$ is a constant cubic root. An orbit shift and scalar phase make $z_t=1$, giving the unitary transition
$6^{-1/2}\left(\begin{smallmatrix}C(a)&C(b)\\C(b)^\dagger&-C(a)^\dagger\end{smallmatrix}\right)$, $C(a)_{rs}=a_{r-s}$. Its flat factorization is, for $r_t=|\alpha_t\beta_t|$,
\begin{equation}
 x_t={\rm i}\alpha_t\beta_t/r_t,\quad y_t=-{\rm i}\bar\alpha_t\beta_t/r_t,\quad
 u_t=\alpha_t+x_t\bar\beta_t,\quad v_t=\alpha_t-x_t\bar\beta_t.
 \label{eq:tc-canonical-lift}
\end{equation}
These entries have modulus one. Solving the two-dimensional factorization equations gives these choices and their signs; each sign change is a common coordinate-pair swap. This is the continuous construction of Sec.~\ref{app:triple-seeds}.

If $A,B$ are dependent, $(|\alpha_t|^2/|\beta_t|^2)(|\alpha_{t+1}|^2/|\beta_{t+1}|^2)$ is constant. Positivity and the amplitude sums force $|\alpha_t|^2=|\beta_t|^2=1/2$, hence $\gamma=0$. The three consecutive ratios are therefore the cubic roots in some order, giving, after scalar normalization,
$a_s=\omega^{\varepsilon s^2+ls}$, $b_s=\omega^{\eta s^2+ms}$, $\varepsilon,\eta=\pm1$, $l,m\in\mathbb Z_3$.
The Gauss sums are $\alpha_t={\rm i}\varepsilon\omega^{-\varepsilon(t+l)^2}/\sqrt2$ and $\beta_t={\rm i}\eta\omega^{-\eta(t+m)^2}/\sqrt2$; dependence requires $\varepsilon=\eta$.
The remaining zero-sum equation has three equal-modulus terms, hence an equilateral triple. Every permutation of cubic roots is affine, giving exactly $w_t=\omega^j$ or $w_t=\omega^j\omega^{\varepsilon t}$. The first returns the continuous family. The second gives $z_t=z_0\omega^{-jt-\varepsilon t(t-1)/2}$. For general unit $z_t$, the lift is explicitly
\begin{equation}
 \begin{gathered}
 x_t={\rm i}\bar z_t\alpha_t\beta_t/r_t,\qquad
 y_t=-{\rm i}\bar\alpha_t\beta_t/r_t,\\
 u_t=\alpha_t+z_tx_t\bar\beta_t,\qquad
 v_t=\alpha_t-z_tx_t\bar\beta_t.
 \end{gathered}
 \label{eq:tc-dependent-lift}
\end{equation}
Thus only the output $x_t$ of Eq.~\eqref{eq:tc-canonical-lift} changes by $z_t^{-1}$; its previously computed $y_t,u_t,v_t$ stay fixed. Quadratic characters cancel in $x_t,y_t$, and $v_t=-{\rm i}u_t$; orbit shifts and scalar phases remove the linear characters, leaving a tensor-product qubit--qutrit triple.

Both signs give one isolated class by unitary equivalence. For $\widehat F_3=F_3/\sqrt3$, $D_\sigma=\diag(\omega^{\sigma t^2})$ and $P_\sigma e_s=e_{\sigma s}$, direct summation gives
\begin{equation}
 \widehat F_3D_\sigma\widehat F_3={\rm i}\sigma D_{-\sigma}\widehat F_3P_\sigma D_{-\sigma},
 \qquad\sigma=\pm1.
 \label{eq:tc-product-equivalence}
\end{equation}
The common unitary $\widehat F_2\otimes\widehat F_3$ and exchange of the first two bases relate the signs up to column monomials. At the product triple the Jacobian of Eq.~\eqref{eq:tc-equations} has rank $55$ on the $55$ dephased variables, so the inverse function theorem isolates its equivalence class. This covers both coefficient cases and all boundary parameters. The basis formulas in Methods Eqs.~\papereqref{eq:methods-isolated-triple-bases} and~\papereqref{eq:methods-explicit-continuous-bases}, and the diagonalization in Sec.~\ref{app:triple-seeds}, use normalized factors $\widehat F_n=F_n/\sqrt n$; transition Hadamards have unit entries.
Common row operations, independent column operations and basis exchange transport the symmetry unitary. Simultaneous conjugation preserves both classified types, and a transpose transfer is basis exchange followed by conjugation. Recognition therefore applies to the original triple.

\subsubsection{The two crossings in the Fourier chart}
\label{sec:tc-crossings}
Translate each exact crossing to $x=0$ in the horizontal slice. The certificate gives its two commuting order-three actions, the orthogonal two-dimensional kernel planes $K_0,K_1$ fixed by the respective actions, a normal inverse $A$ with range $V=(K_0\oplus K_1)^\perp$, and an obstruction map $B$. For $J=DF(0)$ and $\mathcal K=K_0\oplus K_1$, verify
\begin{equation}
 J\mathcal K=0,\qquad BJ=0,\qquad BD^2F(0)[u,v]=u\otimes v\quad(u\in K_0,\ v\in K_1).
 \label{eq:tc-crossing-structure}
\end{equation}
Independent kernel vectors and the inverse bound below give rank $51$. Average $A$ over the nine-element group and replace $B$ by $\frac19\sum_gS_g^{-1}BY_g$, where $S_g$ acts on $K_0\otimes K_1$. This makes both maps equivariant, preserving the identities and norm bounds.
For $\mathcal H=AF$, $\mathcal G=BF$, the bounds used at both crossings are
\begin{equation}
 \begin{gathered}
 \|I_V-AJ|_V\|\le d=10^{-14},\quad
 \|D^2\mathcal H(0)\|\le h_2=1167/500,\quad
 \|D^2\mathcal G(0)\|\le g_2=147/20,\\
 \|D^3\mathcal H\|\le63,\quad\|D^3\mathcal G\|\le105,\\
 \|D^3\mathcal H[\mathbb R^{72},K_0,K_1]\|\le7,\quad
 \|D^3\mathcal G[\mathbb R^{72},K_0,K_1]\|\le12,\\
 \|D^3\mathcal H[\mathbb R^{72},\mathbb R^{72},\mathcal K]\|\le29,\quad
 \|D^3\mathcal G[\mathbb R^{72},\mathbb R^{72},\mathcal K]\|\le47.
 \end{gathered}
 \label{eq:tc-crossing-bounds}
\end{equation}
These are Euclidean multilinear norms, bounded by matrix flattenings. For the $720$ Fourier frequencies, set $D_{ij}=\ell_i\cdot\ell_j$, $U_{ij}=\ell_i\Pi_{K_0}\ell_j^T$, $V_{ij}=\ell_i\Pi_{K_1}\ell_j^T$. Exact summation gives
\begin{equation}
 \max_i\sum_j|D_{ij}|^3=308,\quad
 \max_i\sum_j|D_{ij}U_{ij}V_{ij}|<\tfrac72,\quad
 \max_i\sum_j|D_{ij}^2(U_{ij}+V_{ij})|<61.
 \label{eq:tc-crossing-frequency}
\end{equation}
The coefficient-amplitude matrices after $A,B$ have norms at most $63/\sqrt{308}$ and $105/\sqrt{308}$. Multiplication by the square roots of these row-sum bounds gives the third-derivative bounds; differentiating and summing signed terms gives the center Hessians. Thus finite matrix and frequency checks verify every input, without a local-solution assumption.

Choose $\rho=1/50$, $R=201/10000$, $\eta=1/1000$. The product $\|k\|\le\rho$, $\|y\|\le\eta$, $k\in\mathcal K$, $y\in V$, lies in $\|k+y\|<R$. For $y\mapsto y-\mathcal H(k+y)$,
\begin{equation}
 q=d+h_2R+63R^2/2=0.05963971500001<1,\qquad
 h_2\rho^2/2+63\rho^3/6+q\eta<\eta.
 \label{eq:tc-crossing-contraction}
\end{equation}
Contraction gives a unique analytic equivariant graph $y=\phi(k)$ containing every zero in the product. With $D=1-q$, $r=\|k\|$, comparison with zero gives
$\|\phi(k)\|\le(h_2r^2/2+63r^3/6)/D\le Cr^2$, $C=3/2$, and $\|k+\phi(k)\|\le\chi r$, $\chi=1001/1000$.
Every zero in the full ball $\|x\|\le\rho$ belongs to this product: Taylor's theorem gives $(1-d)\|y\|\le h_2\rho^2/2+63\rho^3/6$, hence $\|y\|<0.000551<\eta$.

For $f(u,v)=\mathcal G(u+v+\phi(u+v))$, equivariance gives $f(u,0)=f(0,v)=0$, since the action fixing each axis has no nonzero fixed vector in the obstruction output. The graph axes need not themselves consist of physical triples. Implicit differentiation gives, with $X=\chi r$,
\begin{equation}
 \begin{aligned}
 t(r)&=(h_2X+29X^2/2)/D,\\
 s(r)&=[h_2+7X+2(h_2+29X)t(r)+(h_2+63X)t(r)^2]/D,\\
 m(r)&=g_2X+X(47r+105Cr^2)/2,\\
 \varepsilon(r)&=12X+2(g_2+47X)t(r)+(g_2+105X)t(r)^2+m(r)s(r).
 \end{aligned}
 \label{eq:tc-crossing-majorants}
\end{equation}
These bound respectively $D\phi$, $D^2\phi[K_0,K_1]$, $D\mathcal G|_V$ on the graph, and the defect of $D^2f[u,v]$ from $u\otimes v$ for unit $u,v$. The last expression is the chain rule: mixed-Hessian change, two single graph-derivative terms, the double graph-derivative term, and $D\mathcal G\,D^2\phi$.
The resulting polynomial $\varepsilon(r)=\sum_{n=1}^8c_nr^n$ has nonnegative rational coefficients. Integrating over the rectangle from $(0,0)$ to $(u,v)$ yields
\begin{equation}
 \|f(u,v)-u\otimes v\|\le\delta\|u\|\|v\|,\qquad
 \delta=\tfrac7{12}c_1\rho+\sum_{n=2}^8\frac{c_n\rho^n}{n+1}<0.885229<1.
 \label{eq:tc-crossing}
\end{equation}
Indeed the first radial moment is at most $\rho/\sqrt3<7\rho/12$; for $n\ge2$, convexity in $\|u\|^2/(\|u\|^2+\|v\|^2)$ bounds it by $\rho^n/(n+1)$. The displayed constants verify $\delta$ by rational arithmetic. Since $\|u\otimes v\|=\|u\|\|v\|$, a zero has $u=0$ or $v=0$; graph equivariance then supplies one of the two physical symmetries, and recognition applies.

In the Fourier core $\|w\|\le r_H/16$, fourteen of the sixteen patterns pass the ordinary symmetry test. The other two have full joint phase-distance bounds below $0.018312<1/50$, after common gauge and integer-winding alignment with the exact crossings. The crossing theorem therefore covers their entire enclosures. For each closed shell cube $\mathcal B$ in the subdivision, the local tests cover $\mathcal B\cap\{\|w\|\le r_H\}$. Checking every shell leaf and every core-containment and outside decision covers the original Fourier ball of radius approximately $0.13$, including its boundary. The separate quartet radius $7/40$ is not used for triple classification.

\subsubsection{Arithmetic and the global conclusion}
\label{sec:tc-verification}
Numerical searches propose centers, multipliers, inverses and frames; their stored binary values are exact rational inputs whose defects are verified. The verifier uses outward-rounded interval arithmetic and, for some comparisons in the initial vector cover, binary64 arithmetic with explicit error bounds. For example, $\|M\|<b$ and $\lambda_{\min}(S)>a$ follow from interval $LDL^T$ factorizations of $b^2I-M^TM$ and $S-aI$ with positive pivots. Cell geometry, windings, permutations and graph enumerations are exact.

The numerical tests in Stage~C discard a vector box or remove an edge only if a strict comparison survives a margin of $10^{-7}$. We use binary64 arithmetic rounded to nearest with gradual underflow. Dot products use binary64 products and sums, possibly fused. The complex modulus uses a scaled sum of squares or a hypot routine with error at most one ulp. For $u=2^{-53}$ and $\eta=2^{-1075}$, a correctly rounded basic operation with finite result and no overflow satisfies
\begin{equation}
 |\operatorname{fl}(z)-z|\le u|z|+\eta .
 \label{eq:tc-underflow}
\end{equation}
We use the outward bounds on the radii from the phase enclosure, including the error in converting the center to cycles. Frame entries and normalized real weights have magnitude at most one, the input radii in cycles are at most one, and the supplied operator-norm bound $K$ is at most three. A finite proposed weight is used as its exact binary value because the support inequalities hold for every such weight. The bin and center exponentials have complex absolute error at most $8u$. 

For each real or complex intermediate $x$, carry bounds $|x|\le A$ and $|\widehat{x}-x|\le E$, where $\widehat{x}$ is its computed value. Write $(B,F)$ for a second input and $\varepsilon=2^{-140}$. Before rounding a sum or difference, its error is at most $E+F$. Before rounding a product, its error is at most
\begin{equation}
 AF+BE+EF .
\end{equation}
For division by a positive real scalar $y\ge d>F$, the error before rounding is at most $(E+AF/d)/(d-F)$. After each of these operations, for an ideal magnitude bound $M$ and the propagated error $e$, use $e+16u(M+e)+\varepsilon$. This covers the real operations and the real-component evaluation of complex products and scalings in these comparisons. For a sum of $n\le36$ terms with bounds $(A,E)$ per term, we use $nE+64nu\,n(A+E)+n\varepsilon$, allowing any addition order. Absolute values of real numbers and selection of minima or maxima do not amplify the largest input error.

For norms we first hold the floating-point inputs fixed. Let $N^2$ be a sum of at most $36$ squared components, including the weighted norm $N^2=\sum_{ij}y_i^2J_{ij}^2$, for which the implementation squares the two factors separately. The computed nonnegative sum differs from $N^2$ by at most $\gamma_{40}N^2+2^{50}\eta$, where $\gamma_{40}=40u/(1-40u)$. Indeed, each summand passes through at most three multiplications and $35$ additions. Each input factor is below $2^{16}$ in magnitude, so an additive error in its square is multiplied by at most $2^{33}$ before accumulation. Summing these additive errors gives the stated allowance. Taking a square root gives error at most $128uN+\varepsilon$, including rounding the square root, since $|\sqrt a-\sqrt b|\le\sqrt{|a-b|}$ for the additive part. For uncertain inputs we add their Euclidean error using the triangle inequality. In the weighted norm each component $y_iJ_{ij}$ uses the product error above. This also bounds norms at zero.

We propagate these bounds through the expressions for $S_i,q_{ji},J_{ij},B_{ik}$ and the remainders in Eqs.~\eqref{eq:local-linear}--\eqref{eq:local-remainder} and~\eqref{eq:local-edges}--\eqref{eq:local-joint}. In particular, the matrix product for $S_i$ sums six complex products before division by $\sqrt6$. In the joint test we bound $J_A^Ty_A+D^Ty_e$, $J_B^Ty_B-D^Ty_e$ and $B_A^Ty_A+B_B^Ty_B$ before taking their respective norms. For each test the table bounds the sum of the absolute errors of the two compared sides. The exact threshold includes the intended decimal margin at the position where the code adds it.
\begin{equation}
 \begin{array}{l|c}
 \text{comparison} & \text{combined absolute error}\\\hline
 \text{single-vector component, squared norm and scalar Taylor tests}
       & 2.1\times10^{-11}\\
 \text{weighted vector test} & 3.1\times10^{-11}\\
 \text{orthogonal and unbiased edge tests} & 2.4\times10^{-12}\\
 \text{joint vector and edge test} & 6.3\times10^{-11}
 \end{array}
 \label{eq:tc-roundoff}
\end{equation}
Every bound is strictly below $10^{-7}$. Thus a computed rejection implies that the exact necessary inequality fails. For label containment we separately check the integer grid indices and the radius. Before nearest-integer conversion, each computed midpoint times $192$ differs from its exact integer by less than $1.9\times10^{-12}$. The wrapped index differences are then exact. Propagating through division by $192$, mean subtraction, the norm and the construction of the inflated radius gives error below $1.5\times10^{-13}$. The stored radius therefore contains every assigned bin because its additive allowance is $10^{-9}$. The native cover repeats the bin assignment with cutoff $1/50$. For an exact centered distance $d$, $6\cdot192^2d^2$ is an integer. Since $6\cdot192^2/50^2=88.4736$ and $d+1/50<2$, the distance is more than $10^{-6}$ from the cutoff, so both evaluations assign the same bins. The Stage~C bounds also apply to pair tests on retained fine bins. Inconclusive tests, arithmetic failures, resource limits and incomplete enumerations remain unresolved.

A certificate contains the original ball's center, frame, radius and reconstruction choice, its complete subdivision, and each non-outside leaf's record, whose containment and branch are checked as in Sec.~\ref{sec:tc-subdivision}. A leaf closed by the tests of the preceding subsections stores nothing else; they are rerun from the record. A normal-form leaf also stores its reference completion, frames and preconditioners and the boxes, parts and tubes used by its normal-form tests. Every bound is recomputed from these exact binary proposals, in Arb or exact arithmetic except for the margined binary64 tests of Stage C bounded above, and symmetry and crossing matrices satisfy the exact identities above. For each ordering discharge, the certificate stores the choice of transition $M$ and the row and column alignment $(a,b)$. The verifier recomputes both score bounds and checks Eq.~\eqref{eq:ordering-discharge} with the same $\Gamma_F$ throughout. A certificate using this discharge is restricted to triples satisfying Eq.~\eqref{eq:minimum-transition}.

\begin{proposition}[Finite triple-classification certificate; computer-assisted]\label{prop:tc-certificate}
Suppose the original atlas covers the non-Tao Hadamard classes, every one of its $2{,}775$ original seed balls has a complete certificate classifying at least the triples satisfying Eq.~\eqref{eq:minimum-transition}, and the entire original Fourier chart has the core and shell certificates just described. Then every MUB triple in $\mathbb C^6$ is equivalent to a member of the continuous construction or the isolated Fourier triple.
\end{proposition}
\begin{proof}
By Lemma~\ref{lem:minimum-transition}, order the triple as $(I_6,H/\sqrt6,K/\sqrt6)$ with $H$ minimizing $\Gamma_F$. If $H$ is in Tao's class there is no completing basis~\cite{BrierleyWeigert2009}; otherwise the atlas, closed subdivisions, branch identifications and column covers retain $H,K$. A zero satisfies neither a strict exclusion nor a strict ordering discharge. Symmetry or crossing and recognition therefore identify the triple, proving the proposition and Conjecture~1 under its finite certification condition.
\end{proof}

The verification reports all original seed balls closed, four with minimum-transition scope (ball IDs $1035$, $1217$, $1622$ and $2705$), and a separate complete Fourier-chart calculation.

\begin{proof}[Proof of Theorem~\paperref{thm:all-mubs}]
Up to equivalence every set of MUBs contains $I_6$, and every pair is $\{I_6,H/\sqrt6\}$ with $H$ complex Hadamard; by Theorem~\paperref{thm:reconstruction}, $H$ is Tao's matrix, $F_6$ or $H_+(\theta)$ for a strict seed $\theta$. Proposition~\ref{prop:tc-certificate} classifies the triples. Both types of triple contain a pair whose transition lies in the Fourier family, and no such pair extends to four MUBs~\cite{Jaming2009}. Hence no set contains four bases; Sec.~\ref{sec:vector-exclusion} gives a second, direct exclusion of quartets.
\end{proof}

\subsubsection{Seeds of all mutually unbiased triples}
\label{app:triple-seeds}
Methods Eqs.~\papereqref{eq:methods-isolated-triple-bases} and~\papereqref{eq:methods-explicit-continuous-bases} give the isolated Fourier triple and continuous family, with $\omega=e^{2\pi {\rm i}/3}$. We now specify the diagonal matrices $D,R,S$ and the four-phase seeds. All phases are modulo $2\pi$. For $\sqrt6U_2^\dagger U_3$, take the upper-left corner; for $\sqrt6U_3^\dagger$, take rows $(1,4,2)$ and columns $(1,2,3)$, dephasing before reading the seed.

The isolated triple tensors corresponding qubit and qutrit MUB bases; all three transition Hadamards are equivalent to $F_6$. Its two specified seeds are $(2\pi/3,-2\pi/3,-2\pi/3,2\pi/3)$ and $(0,0,-2\pi/3,2\pi/3)$, respectively; conjugation negates both.
For the continuous family start from $(\theta_1,\theta_2,\theta_1-\theta_2,\theta_1)$ and, for $\ell=1,2,3$, set
\begin{equation}
 \beta=\frac{\theta_2-2\theta_1+2\pi(\ell-1)}3,\qquad
 \gamma=-\frac{\theta_1+\theta_2+2\pi(\ell-1)}3,\qquad
 \sigma=e^{-{\rm i}\beta}+e^{{\rm i}(\beta-\gamma)}+e^{{\rm i}\gamma}.
\end{equation}
The allowed parameter pairs are exactly those satisfying
\begin{equation}
 |\sigma|^4+18|\sigma|^2+8\Real(\sigma^3)-27\le0.
 \label{eq:triple-parameter-domain}
\end{equation}
Then the roots $\rho_1,\rho_2,\rho_3$ of $t^3+\sigma t^2-\bar\sigma t-1=0$, counted with multiplicity, have unit modulus. For every ordering, form
\begin{equation}
 A=\begin{pmatrix}1&e^{{\rm i}\beta}&e^{{\rm i}\gamma}\\e^{{\rm i}\gamma}&1&e^{{\rm i}\beta}\\e^{{\rm i}\beta}&e^{{\rm i}\gamma}&1\end{pmatrix},\quad
 B=\begin{pmatrix}1&\rho_1^{-1}&\rho_3\\\rho_3&1&\rho_1^{-1}\\\rho_1^{-1}&\rho_3&1\end{pmatrix},\quad
 H=\begin{pmatrix}A&B\\B^\dagger&-A^\dagger\end{pmatrix}.
 \label{eq:triple-blocks-from-seed}
\end{equation}
These functions of $\theta_1,\theta_2$ give a Hadamard matrix. With the normalized Fourier factors specified in Sec.~\ref{sec:tc-recognition}, compute
\begin{equation}
 \begin{aligned}
 \lambda_k&=(\widehat F_3A\widehat F_3^\dagger)_{kk}/\sqrt6,&
 \mu_k&=(\widehat F_3B\widehat F_3^\dagger)_{kk}/\sqrt6,\\
 D_{kk}&={\rm i}\epsilon_k\lambda_k\mu_k/|\lambda_k\mu_k|,&
 R_{kk}&=\lambda_k+D_{kk}\bar\mu_k,\quad S_{kk}=\lambda_k-D_{kk}\bar\mu_k,
 \end{aligned}\qquad k=1,2,3,\quad\epsilon_k=\pm1.
 \label{eq:triple-diagonal-factors}
\end{equation}
The denominators are nonzero and the diagonal entries have unit modulus throughout the allowed domain. Substitution in Methods Eq.~\papereqref{eq:methods-explicit-continuous-bases} gives $\sqrt6U_2^\dagger U_3=H$, whose upper-left corner has the prescribed seed. The second specified corner has seed $(\theta'_1,\theta'_2,-2\pi/3,2\pi/3)$, where
\begin{equation}
 \theta'_1=\arg\frac{R_{22}S_{22}D_{11}}{R_{11}S_{11}D_{22}},\qquad
 \theta'_2=\arg\frac{R_{33}S_{33}D_{11}}{R_{11}S_{11}D_{33}}.
\end{equation}
All root orderings, independent signs and repeated-root boundary cases are retained; coincident seed coordinates need not represent equivalent triples.

\subsection{Fourier vanishing}
\label{sec:fourier-vanishing}
This section proves the Fourier vanishing conjecture of Matolcsi, Ruzsa and Weiner from the reconstruction and linear algebra. Sec.~\ref{sec:fv-identities} derives identities between complementary blocks, and Sec.~\ref{sec:fv-moments} uses them to make every row and column moment vanish.

For an order-six Hadamard $H$ and a three-element row set $I$, define~\cite[Sec.~3]{MaxwellBrierley2015}
\begin{equation}
 M_I(H)=\frac16\sum_{j=1}^{6}\prod_{i\in I}H_{ij}\prod_{k\notin I}\overline{H_{kj}}.
 \label{eq:fourier-moment}
\end{equation}
Vanishing means that the entrywise products of complementary row triples are orthogonal. Column phases cancel and row phases give a common unit factor, so simultaneous vanishing respects equivalence.

\begin{proposition}[Fourier vanishing]\label{prop:fourier-vanishing}
If $H$ is not equivalent to Tao's matrix, then $M_I(H)=M_I(H^T)=0$ for every three-element set $I$.
\end{proposition}
For Tao's symmetric representative $S_6$~\cite[Eq.~(10)]{wuttig2026completeclassificationcomplexhadamard}, three summands are $1$ and three $\omega^2$, so $M_{\{1,2,3\}}(S_6)=(1+\omega^2)/2=-\omega/2\ne0$. We prove the conjecture~\cite[Conjecture~2.3]{MatolcsiRuzsaWeiner2013} from reconstruction and linear algebra, independently of the numerical atlas.

\subsubsection{Complementary-block identities}
\label{sec:fv-identities}
For $H=\left(\begin{smallmatrix}E&B\\C&D\end{smallmatrix}\right)$, $B,C$ have the same $g,\eta,\chi$. Their $\Psi$ values vanish if $g=0$, coincide and are real if $g>0,R=0$, and are conjugate if $R>0$ by Proposition~\ref{prop:nonmatched}. Exchanging column halves gives, in every case,
\begin{equation}
 \Psi(H[I,J])=\overline{\Psi(H[I^c,J^c])},\qquad |I|=|J|=3.
 \label{eq:conjugate-corners}
\end{equation}
Write $X^{[2]}$ for entrywise squaring and $P_X=\prod_{a,b}X_{ab}$. Cayley's identity~\cite[Theorem~1.2]{Singer2004Borchardt} gives, for any $3\times3$ matrix whose entries have magnitude one,
\begin{equation}
 \det X^{[2]}=\perm(X)\det X-2P_X\overline{\det X},\qquad
 \Psi(X)=2|\det X|^2+\frac{\det X\det X^{[2]}}{P_X}.
 \label{eq:square-identity}
\end{equation}
The even and odd permanent terms each have product $P_X$, and their pairwise-product sums are $P_X$ times their conjugate sums. Subtract their sums of squares, using $\sum a^2=(\sum a)^2-2\sum_{i<j}a_ia_j$, for the first formula; multiply by $\det X/P_X$ for the second. A singular block therefore has a singular entrywise square.

Define $(A_I)_{ij}=H_{ij}^2/\prod_{r\in I}H_{rj}$, $i\in I$, write $A_{I,J}$ for its columns in $J$, and set $\sigma(I,J)=\sum_{i\in I}i+\sum_{j\in J}j$.
Jacobi's identity with $H^{-1}=H^\dagger/6$, then Eqs.~\eqref{eq:conjugate-corners}--\eqref{eq:square-identity}, give
\begin{equation}
\begin{aligned}
 \det H[I^c,J^c]&=(-1)^{\sigma(I,J)}\frac{\det H}{216}\,\overline{\det H[I,J]},\\
 \det A_{I,J}&=(-1)^{\sigma(I,J)}\frac{\overline{\det H}}{216}\,
                  \overline{\det A_{I^c,J^c}}.
\end{aligned}
 \label{eq:fv-complementary-minors}
\end{equation}
For $\det H[I,J]\ne0$, cancel this factor after substitution. Otherwise Jacobi's identity and Eq.~\eqref{eq:square-identity} make both complementary minors of the entrywise squares zero, proving the same relation.

\subsubsection{Vanishing of the row and column moments}
\label{sec:fv-moments}
If $A_I$ has rank three, Eq.~\eqref{eq:fv-complementary-minors} makes $\overline{A_{I^c}}$ rank three. Permute columns to make the first three of $A_I$ nonsingular; the complementary three of $\overline{A_{I^c}}$ are then nonsingular. Invertible row changes give $A=(I_3\ U)$ and $C=(V\ I_3)$. At $J_0=\{1,2,3\}$ both pivot minors equal one, so $\det A_J=(-1)^{\sum_{j\in J}j-6}\det C_{J^c}$. Exchanging columns $1,4$ gives $U_{11}=-V_{11}$, and the nine exchanges give $V=-U^T$. Undoing the row changes yields
\begin{equation}
 A_IA_{I^c}^\dagger=0,\qquad
 (A_IA_{I^c}^\dagger)_{ik}
 =6M_{\{i,\ell,m\}}(H),\quad i\in I,\quad I^c=\{k,\ell,m\}.
 \label{eq:fv-pairing}
\end{equation}
These nine moments and their conjugates $M_{S^c}=\overline{M_S}$ give all eighteen mixed row triples; only $M_I,M_{I^c}$ remain.

A nonsingular $H^{[2]}[I,J]$ supplies such a rank-three $A_I$. Its rows form a basis, so any nonzero row $H^{[2]}[k,J]$, $k\notin I$, can replace some row $i\in I$ while preserving nonsingularity. For the resulting triple $I'$, Eq.~\eqref{eq:fv-pairing} using rows $k,i$ gives $M_{I^c}$; its conjugate gives $M_I$.

If $H$ is not equivalent to $F_6$, Proposition~\ref{prop:critical} gives a strict corner. Its block $B$ has nonreal $\Psi(B)$, so Eq.~\eqref{eq:square-identity} forces $\det B^{[2]}\ne0$. For the Fourier matrix we obtain directly
\begin{equation}
 6M_I(F_6)=\sum_{j=1}^6e^{2\pi {\rm i}(2\sum_{i\in I}i-21)(j-1)/6}=0,
\end{equation}
because $2\sum_{i\in I}i-21$ is odd and therefore not divisible by six. Finally, apply the row result to $H^T$. Transposition preserves equivalence and Tao's representative is symmetric, so $H^T$ is non-Tao whenever $H$ is. This proves the column moments as well, and with them Proposition~\ref{prop:fourier-vanishing}.

\begingroup
\setlength{\abovedisplayskip}{6pt plus 1pt minus 1pt}
\setlength{\belowdisplayskip}{6pt plus 1pt minus 1pt}
\setlength{\abovedisplayshortskip}{3pt plus 1pt}
\setlength{\belowdisplayshortskip}{3pt plus 1pt minus 1pt}
\setlength{\textfloatsep}{10pt plus 1pt minus 1pt}
\setlength{\intextsep}{8pt plus 1pt minus 1pt}

\subsection{Mutually unbiased constellations}
\label{sec:constellation-certificates}
Each constellation contains $\{\mathbf e_1,\ldots,\mathbf e_5\}$ from the identity basis. The other vectors are listed without their normalization factor $1/\sqrt6$. Put $\omega=(-1+{\rm i}\sqrt3)/2$, $\mathbf1=(1,1,1,1,1,1)^T$, and define
\begin{equation}
 \begin{aligned}
 \Lambda&=\diag(1,1,\omega,\bar\omega,\omega,\bar\omega),\qquad
 \mathbf h=(1,-1,T,Q,-T,-Q)^T,\\
 \mathbf x'&=(\bar x_2,-1,T\bar\omega\bar x_5,Q\omega\bar x_6,
              -T\bar\omega\bar x_3,-Q\omega\bar x_4)^T,
 \qquad \mathbf x=\mathbf b,\mathbf c.
 \end{aligned}
 \label{eq:constellation-vectors}
\end{equation}
The three additional groups are
\begin{equation}
 \begin{aligned}
 \{5,3,3,3\}_6:\quad&
 \{\mathbf1,\Lambda\mathbf1,\Lambda^2\mathbf1\},\quad
 \{\mathbf b,\Lambda\mathbf b,\Lambda^2\mathbf b\},\quad
 \{\mathbf c,\Lambda\mathbf c,\Lambda^2\mathbf c\},\\
 \{5,4,3,1\}_6:\quad&
 \{\mathbf1,\mathbf h,\Lambda^2\mathbf h,\Lambda\mathbf1\},\quad
 \{\mathbf b,\Lambda\mathbf b,\Lambda^2\mathbf b\},\quad\{\mathbf c\},\\
 \{5,5,2,2\}_6:\quad&
 \{\mathbf1,\mathbf h,\Lambda\mathbf h,\Lambda^2\mathbf h,\Lambda^2\mathbf1\},\quad
 \{\mathbf b,\mathbf b'\},\quad\{\mathbf c,\mathbf c'\}.
 \end{aligned}
 \label{eq:constellation-vector-lists}
\end{equation}
The entries are specified separately for each case.

\paragraph*{$\{5,3,3,3\}_6$.}
Take
\begin{equation}
 a=\frac{1+{\rm i}\sqrt{15}}4,\qquad
 \mathbf b=(1,\bar\omega,1,a,\omega,a)^T,\qquad
 \mathbf c=(1,\bar\omega,\bar a,\omega,\bar a,1)^T.
 \label{eq:5333-explicit-seeds}
\end{equation}

\paragraph*{$\{5,4,3,1\}_6$.}
Take $T=Q=1$ and
\begin{equation}
 \mathbf b=(1,-{\rm i},\omega,1,-{\rm i}\omega,-{\rm i})^T,\qquad
 \mathbf c=(1,e,p,pq,{\rm i}p\overline{de},dpq)^T.
 \label{eq:5431-explicit-vectors}
\end{equation}
The four fixed phases are $d=F(\beta)$, $e=F(\bar\omega\beta)$, $p=G(\beta)$ and $q=G(\bar\omega\beta)$, where
\begin{equation}
 \begin{aligned}
 R&=\sqrt{\frac{\sqrt5-1}{2}},\qquad
 \kappa=\sqrt3\,R^3,\qquad
 \eta=\frac{R-1+{\rm i}R^2}{2R^2},\qquad
 \beta^3=\frac{R^2}{4}(-1+{\rm i}R^3),\\
 Y(t)&=2R\operatorname{Im}t,\qquad
 F(t)=-\frac{(Y(t)+1)(2Y(t)-1)}{\kappa}+{\rm i}Y(t),\qquad
 G(t)=\frac{(1-{\rm i})t}{R}\frac{\overline{t+\eta}}{t+\eta}.
 \end{aligned}
 \label{eq:5431-four-phases}
\end{equation}
The real square roots are positive; any cube root $\beta$ gives a constellation.

\paragraph*{$\{5,5,2,2\}_6$.}
Write $\mathbf b=(1,b_2,\ldots,b_6)^T$, $\mathbf c=(1,c_2,\ldots,c_6)^T$ and set
\begin{equation}
 (T,Q,b_2,\ldots,b_6,c_2,\ldots,c_6)
 =\left(\frac{1+{\rm i}t_j}{1-{\rm i}t_j}\right)_{j=1}^{12}.
 \label{eq:constellation-algebraic-entries}
\end{equation}
The real vector $t$ is the unique solution of
\begin{equation}
 \begin{aligned}
 |\mathbf z^\dagger\mathbf b|^2&=|\mathbf z^\dagger\mathbf c|^2=6
 &&\bigl(\mathbf z\in\{\mathbf1,\mathbf h,\Lambda\mathbf h,\Lambda^2\mathbf h,\Lambda^2\mathbf1\}\bigr),\\
 |\mathbf b^\dagger\mathbf c|^2&=|\mathbf b^\dagger\mathbf c'|^2=6
 \end{aligned}
 \label{eq:constellation-cross}
\end{equation}
in the rational box
\begin{equation}
 \|t-m/10^7\|_\infty\le10^{-6},\qquad
 \begin{aligned}
 m=(&88191,-12383399,-17366867,-6227352,-15603052,-1112587,\\
     &34985264,13037208,17392983,1296019,-22919921,7181585)^T.
 \end{aligned}
 \label{eq:5522-isolating-box}
\end{equation}
We first find an approximate solution numerically, then use interval arithmetic to prove that the stated box contains exactly one solution of the overlap equations, which specifies the vector entries as exact algebraic numbers. The remaining overlaps follow from the antiunitary map $J\mathbf x=(\bar x_2,-\bar x_1,T\bar\omega\bar x_5,Q\omega\bar x_6,-T\bar\omega\bar x_3,-Q\omega\bar x_4)^T$, for which $\mathbf x'=J\mathbf x$ when $x_1=1$. Since $J^2=-I$ and $\langle J\mathbf a,J\mathbf b\rangle=\langle\mathbf b,\mathbf a\rangle$, $\langle\mathbf x,J\mathbf x\rangle=0$, so $\mathbf b\perp\mathbf b'$ and $\mathbf c\perp\mathbf c'$. $J$ permutes, up to phases, the rays of the orthonormal basis completing the five-vector group, so the first line of Eq.~\eqref{eq:constellation-cross} carries over to $\mathbf b',\mathbf c'$, the sixth vector by Parseval. Finally $|\langle J\mathbf b,J\mathbf c\rangle|=|\langle\mathbf b,\mathbf c\rangle|$ and $|\langle J\mathbf b,\mathbf c\rangle|=|\langle\mathbf b,J\mathbf c\rangle|$ give the remaining cross overlaps.

\phantomsection\label{sec:strong-unextendibility}
\endgroup
\paragraph*{Strong unextendibility.}
Every classified triple is strongly unextendible: no unit vector is unbiased to all three bases. For the continuous normal form this is Ref.~\cite[Corollary~9.1]{Sun2026Completion}; for the isolated Fourier triple, equivalent to a product triple, it is Ref.~\cite{McNultyWeigert2012Triples}. This resolves Problem~10.3 of Ref.~\cite{McNultyWeigert2026}.
In a hypothetical $\{5,5,5,1\}_6$, complete each five-vector set by its unique orthogonal ray. A unit vector unbiased to the first five members is also unbiased to the sixth, since $|\langle b_6|v\rangle|^2=1-\sum_{r=1}^5|\langle b_r|v\rangle|^2=1/6$. Applying this successively gives three MUBs and a vector unbiased to all three, contradicting strong unextendibility.

\NoFourMUBsSupplementEnd

\end{document}